\documentclass[journal=nalefd,manuscript=article,doi=true]{achemso}
\setkeys{acs}{doi = true}

\usepackage{upgreek}
\usepackage{graphicx}% Include figure files
\usepackage{dcolumn}% Align table columns on decimal point
\usepackage{bm}% bold math
\usepackage{amssymb}
\usepackage{epstopdf}
\usepackage{textcomp}
\usepackage{enumerate}
\usepackage{epsfig,amsmath}
\usepackage[colorlinks]{hyperref}
\usepackage{color}
\usepackage[usenames,dvipsnames]{xcolor}
\usepackage{float}
\usepackage{wasysym}
\usepackage{pifont}
\usepackage{tikz}
\usepackage{array}
\usepackage{colortbl}

\usepackage{MnSymbol}
\usepackage{tabularx}
\usepackage{soul}
\usepackage[utf8]{inputenc}
\usepackage{subcaption}

\newcommand{\be}{\begin{equation}}
\newcommand{\ee}{\end{equation}}
\newcommand{\doi}[1]{\href{http://dx.doi.org/#1}{\nolinkurl{#1}}} % for use with hyperref package to get DOI links to work in the references
\title{Electrical low-frequency $1/f^{\gamma}$ noise due to surface diffusion of scatterers on an ultra low noise graphene platform}
\author{Masahiro Kamada}
\affiliation{Low Temperature Laboratory, Department of Applied Physics, Aalto University School of Science, P.O. Box 15100, 00076 Aalto, Finland}

\author{Antti Laitinen}
\affiliation{Low Temperature Laboratory, Department of Applied Physics, Aalto University School of Science, P.O. Box 15100, 00076 Aalto, Finland}

\author{Weijun Zeng}
\affiliation{Low Temperature Laboratory, Department of Applied Physics, Aalto University School of Science, P.O. Box 15100, 00076 Aalto, Finland}

\author{Marco Will}
\affiliation{Low Temperature Laboratory, Department of Applied Physics, Aalto University School of Science, P.O. Box 15100, 00076 Aalto, Finland}

\author{Jayanta Sarkar}
\affiliation{Low Temperature Laboratory, Department of Applied Physics, Aalto University School of Science, P.O. Box 15100, 00076 Aalto, Finland}

\author{Kirsi Tappura}
\affiliation{Quantum systems, VTT Technical Research Centre of Finland Ltd., P.O. Box 1000, 02044 VTT, Finland}

\author{Heikki Sepp\"a}
\affiliation{Quantum systems, VTT Technical Research Centre of Finland Ltd., P.O. Box 1000, 02044 VTT, Finland}

\author{Pertti Hakonen}
\affiliation{Low Temperature Laboratory, Department of Applied Physics, Aalto University School of Science, P.O. Box 15100, 00076 Aalto, Finland}
\alsoaffiliation{QTF Centre of Excellence, Department of Applied Physics, Aalto University School of Science, P.O. Box 15100, 00076 Aalto, Finland}
\email{pertti.hakonen@aalto.fi}

\date{\today}% It is always \today, today,
\begin{document}

\begin{abstract}
Low-frequency $1/f^{\gamma}$ noise is ubiquitous, even in high-end electronic devices. For qubits such noise results in decrease of their coherence times. Recently, it was found that adsorbed O$_2$ molecules provide the dominant contribution to flux noise in superconducting quantum interference devices. To clarify the basic principles of such adsorbant noise, we have investigated the formation of low-frequency noise while the mobility of surface adsorbants is varied by temperature. In our experiments, we measured low-frequency current noise in suspended monolayer graphene samples under the influence of adsorbed Ne atoms. Owing to the extremely small intrinsic noise of graphene in suspended Corbino geometry, we could resolve a combination of $1/f^{\gamma}$ and Lorentzian noise spectra induced by the presence of Ne. We find that the $1/f^{\gamma}$ noise is caused by surface diffusion of Ne atoms and by temporary formation of few-Ne-atom clusters. Our results support the idea that clustering dynamics of defects is relevant for understanding of $1/f$ noise in general metallic systems.
\end{abstract}

\maketitle 
\vspace{30pt}
Quantum devices in nanotechnology are plagued by $1/f^{\gamma}$ noise. Monolayer graphene devices are no exception, even though they have been found to exhibit ultra low noise \cite{Balandin2013a}. Suspended graphene have been found to provide the lowest noise since they can be made nearly perfectly clean without the influence of defects in the nearby substrate, and they display an ultra-high mobility \cite{Kumar2015a}. Several physical mechanisms have been suggested as the origin of the $1/f^{\gamma}$ noise in graphene, either via fluctuations of the chemical potential or directly via mobility fluctuations  \cite{Heller2010,Pal2011b,Zhang2011a,Kaverzin2012,Pellegrini2013,Arnold2016}. In addition, contact noise has been found to be relevant in many cases \cite{Liu2012a,ZahidHossain2013,Kumar2015a}, which may result from current crowding at the contacts \cite{Toriumi2010,Grosse2011,Karnatak2016}.

In general, the \emph{ad hoc} models considered for graphene can be criticized as they do not arise from a unified concept. The same issue, however, does pertain to various areas of $1/f$ noise \cite{Dutta1981,Hooge1994,Kogan2008,Grasser2020}.  In this work, our goal is to test fundamental aspects of theories based on mobile impurities \cite{Martin1972,Nagaev1982,Robinson1983,Pelz1987,Giordano1989,Klonais1993}. We generate $1/f^{\gamma}$ noise by adsorbing neon atoms onto a graphene membrane. 
We are employing suspended graphene as a platform for studying impurity-induced low-frequency noise, because the inherent $1/f^{\gamma}$ noise level is exceedingly small in mechanically exfoliated, suspended graphene. 
Since the background impurity scattering is almost non-existent in graphene, even weak scatterers such as neon atoms may make a difference in the impurity scattering, and thereby alter the noise substantially.

All solid materials contain structural defects which may diffuse around at room temperature. When  temperature  is  lowered the  diffusion  slows  down  but  it  remains  still  visible  all  the  way down to the quantum tunneling regime.  We are interested in diffusion of defects or impurities and their possible role  in  scattering fluctuations due to clustering  of defects/impurities. Diffusion influences  the  relative  locations of  impurities  which  then  affect  the  total  scattering cross section (length in two dimensions) experienced  by  the  charge carriers traversing the sample.  Variation of the scattering will, in turn, lead to modification of resistance and the fundamental question is whether this will lead to  $1/f^{\gamma}$ noise spectrum with $\gamma \simeq 1$. The answer to this question bears significance also to quantum technology as the interfacial states and adsorbant atoms present important noise sources for qubits \cite{Kumar2016,Lisenfeld2019}.  

Owing to its large surface-to-volume ratio, graphene is always very susceptible to surface adsorbants, and even individual Hall resistance steps have been demonstrated due to single atom adsorption events \cite{Schedin2007}. It has experimentally been demonstrated that adsorbed gas molecules at room temperature will lead to Lorentzian noise spectra, with  characteristics specific to the adsorbent species \cite{Rumyantsev2012}. Noble gases on graphite interact quite weakly with the substrate. At low temperatures, nevertheless, even they leave their specific fingerprints on the low frequency noise. In addition, the surface diffusion of light noble gases remains sufficiently intense at cryogenic temperatures which allows studies of clustering dynamics of such atoms and their influence on the low frequency noise. 

The adsorption of neon on graphene is expected to become strong around 15 Kelvin, the bulk solidification temperature of Ne. Therefore, our experiments were carried out at temperatures from 4 K to 37 K over which we may vary the behavior of neon on graphene from surface diffusion by nearly pure quantum tunneling to thermally driven adsorption-desorption regime. The most interesting regime is attained around $10$ Kelvin where surface diffusion leads to fluctuations in the number and in the distribution of the atoms on the graphene membrane, while desorption rate of Ne can be neglected. In this regime, we may test universal $1/f$ theories based on mobile impurities. Our results can also be employed to determine the adsorption energy of a Ne atom onto a graphene monolayer.

In our experiments, we measured both rectangular suspended graphene flakes and Corbino disks. For suspension of the membranes, we employed HF etching of SiO$_2$ as well as dissolving a LOR sacrificial layer. Our sample fabrication techniques and illustrations of samples can be found in Refs. \citenum{Kumar2018,Laitinen2016}. 
Our most extensive data set was taken using a Corbino disk with distance $L=1.3$ $\upmu$m between the electrodes (inner and outer diameters of 1.8 and 4.5 $\upmu$m, respectively). The value of the gate capacitance $C_g = 1.5 \times 10^{-5}$ F/m$^2$ was determined  using the spread of a Landau level fan diagram \cite{Kumar2018}.

Following the initial characterization at room temperature, the samples were mounted on a Bluefors LD250 dry dilution refrigerator and cooled down to $T$ = $10$ mK. Prior to low-frequency noise measurements, the graphene devices were current annealed at $4\,$K, at which cryopumping guaranteed a UHV level cryogenic vacuum. Cryopumping by the cryostat worked also at higher temperatures against impurity gases such as O$_2$ and N$_2$. In addition, current annealing was employed at regular intervals to clean neon away from the graphene surface, thereby verifying that the results indeed were caused by Ne atoms. The applied gate voltage $V_g$ was employed to determine the charge carrier density according to $n = (V_g-V_g^D) C_g/e$, where $V_g^D$ denotes the gate voltage value of the Dirac point. Measurements of zero-bias resistance $R_0(V_g)$ (see the SI \cite{SI}) yielded the field-effect mobility $\mu_{FE} \sim 10^5$ cm$^2$/Vs using $\mu_{FE} = (\sigma-\sigma_0)/ne$, where the minimum conductivity $\sigma_0$ corresponds to the maximum of measured resistivity at the Dirac point. According to  $\mu_{FE}$ determination, adsorption of Ne increased the mobility of graphene by $\sim 30$\% at small charge densities $|n|\lesssim 2 \times 10^{14}$\,m$^{-2}$. This indicates additional screening of Coulomb impurities by adsorbed Ne. The effect of Ne-induced charge screening/charge modulation could be removed by thermal annealing, which was repeatedly done during our experiments.

The low frequency noise was measured under voltage bias up to $V_b=13$ mV. A transimpedance amplifier (SR570, gain $10^5$) was employed to track the current and the time trace of fluctuations was Fourier transformed using an SRS 785 FFT signal analyzer. The obtained $1/f^{\gamma}$ noise spectra, averaged for 180 s, consisted of 200 FFT points, spanning typically the frequency range of $1 - 200$ Hz. The $V_b^2$ voltage dependence of the noise was verified over the whole employed frequency range. Dependence of the scaled low frequency noise power $S_I/I^2$ on various parameters was typically measured at 10 Hz. 
The intrinsic $1/f$ noise in our Corbino devices is quite low compared to all reported suspended graphene based devices of similar size, including our own earlier results on suspended bilayer graphene \cite{Kumar2015a}. 

\begin{figure}[t!]
	\centering
	\includegraphics[width=.45\linewidth]{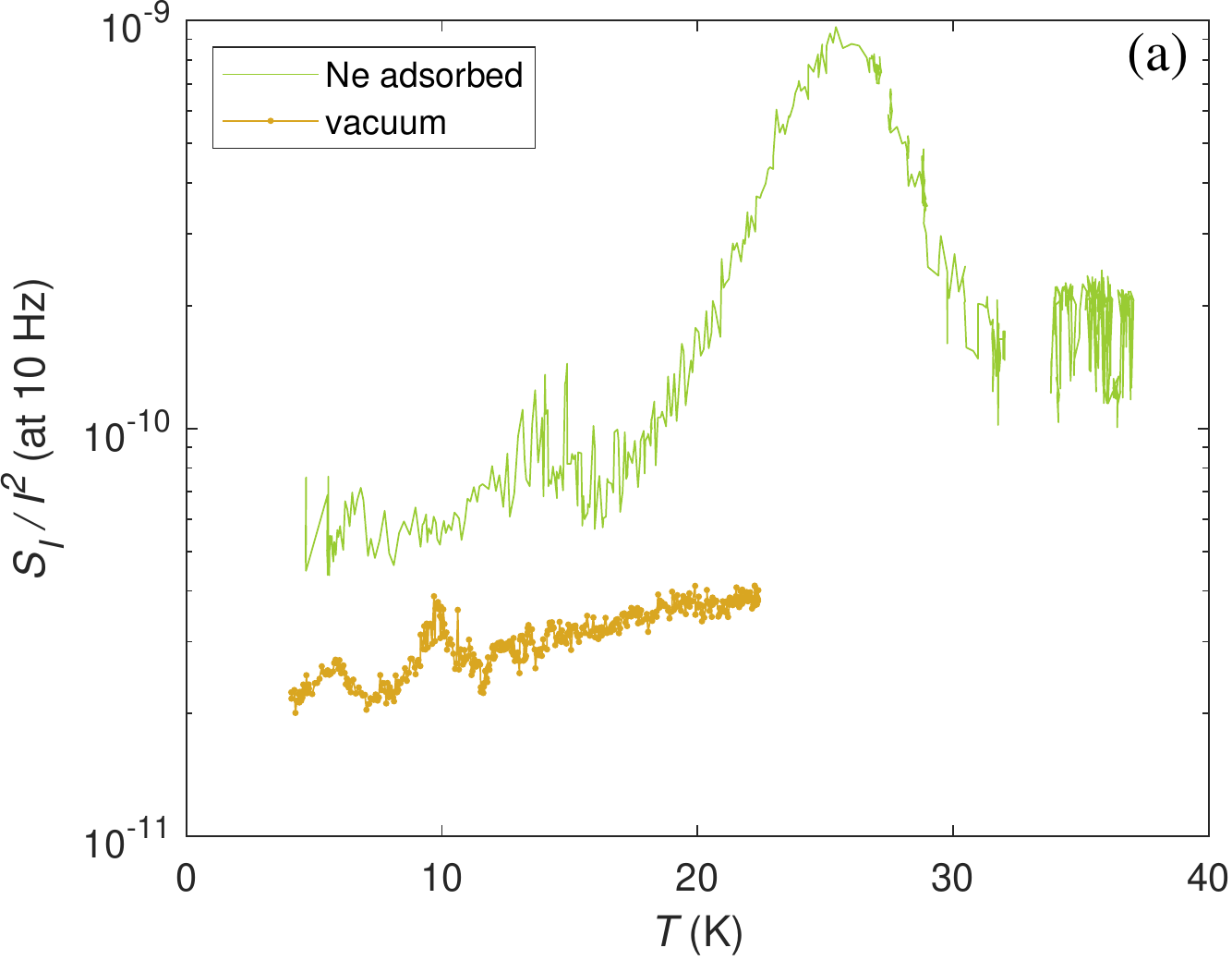}
	\includegraphics[width=.45\linewidth]{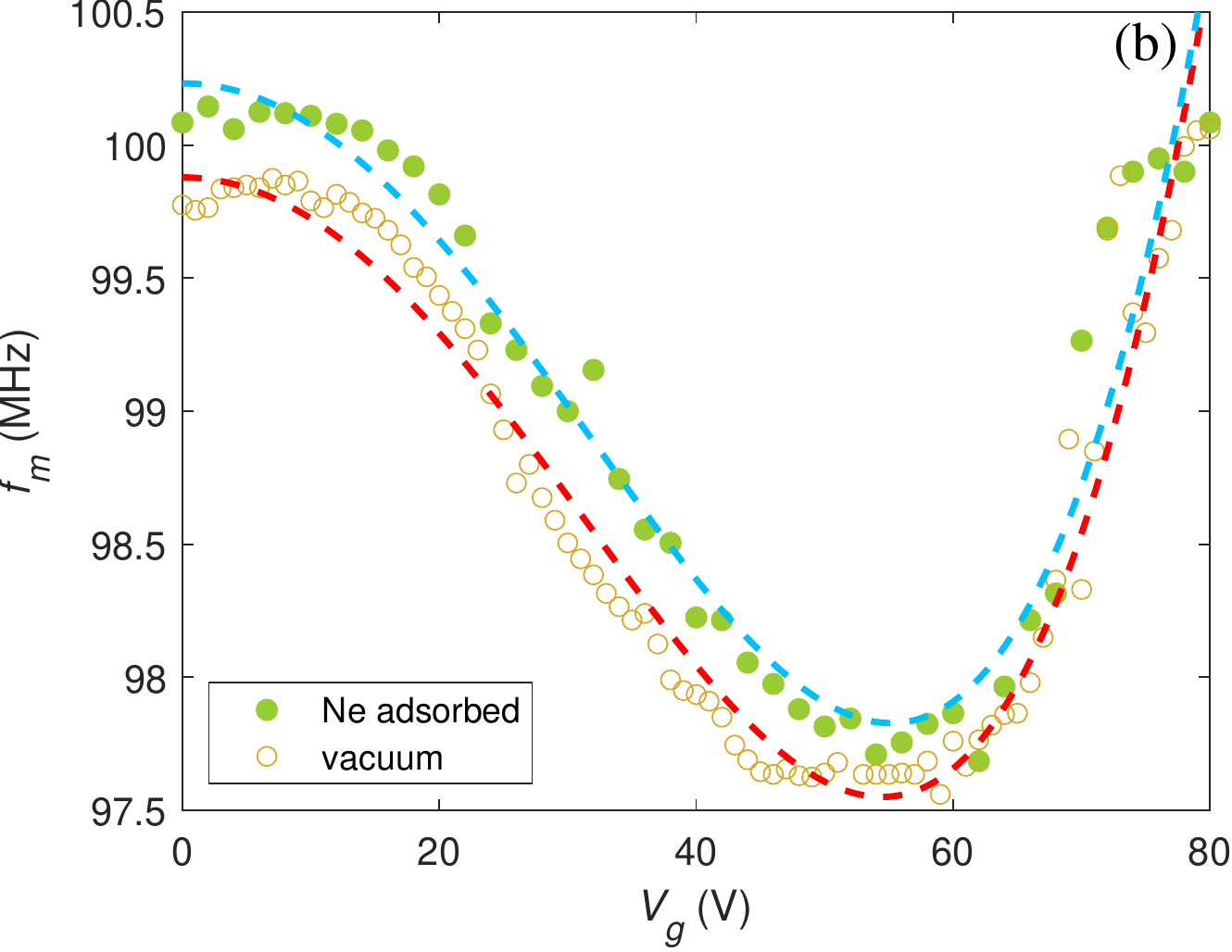}

	\caption{a) Scaled current noise $S_I/I^2$ at 10 Hz as a function of $T$ measured with (upper trace, green) and without (lower trace, orange) adsorbed Ne. 
	b) Mechanical resonance frequency $f_m$ vs. gate voltage $V_g$ measured for Corbino disk with (\textcolor{green}{$\bullet$}) and without (\textcolor{orange}{$\circ$}) adsorbed neon. The re-entrant looking $f_m(V_g)$ (half of W-shape) indicates a crossover from the initial-tension-dominated gate dependence to induced-tension-dominated behavior. The dashed curves are a quartic fits to the data with (upper) and without neon (lower); the shift of the W-shape minimum to larger $V_g$ is also a sign of enhanced tension by neon. \label{NoiseandR}}
\end{figure}

Fig. \ref{NoiseandR} illustrates the basic influence of neon gas on our graphene samples. Fig. \ref{NoiseandR}a displays the measured low-frequency noise at 10 Hz as a function of temperature with and without neon (upper and lower trace, respectively). Without adsorbed Ne, the low-frequency noise increases approximately linearly with $T$, whereas the background level of $S_I/I^2$ with Ne is nearly constant up to $T \simeq 20$ K. At temperatures $25 - 30$ K, there is a strong peak in the noise which can be attributed to the dynamics of neon atoms on the graphene surface, governed basically by adsorption/desorption processes; the $S_I/I^2$ reading at 10 Hz with Ne in Fig. \ref{NoiseandR}a corresponds to averaged noise obtained by fitting of a linear combination of $1/f^{\gamma}$ and a Lorentzian spectrum to the data over $1-100$ Hz. The smaller peak at $T=12-16$ K may be due to residual hydrogen present in the system, but more likely it is related to diffusion phenomena of neon atoms across the graphene sample. The quality of the sample is exemplified by the product $f \times S_I/I^2 \simeq 2 \times 10^{-10}$, which is on par with the best noise levels reported so far \cite{Kumar2015a}.

Fig. \ref{NoiseandR}b displays the mechanical resonance frequency measured for the graphene Corbino disk that was employed for the majority of our noise experiments. In fact, this device displayed several weak mechanical resonances with slightly different frequency, which we interpret as splitting of the fundamental mode into several local resonances due to non-uniform strain in the membrane. As seen in Fig. \ref{NoiseandR}b, the displayed resonant frequency is increased by 2\textperthousand\ in the presence of adsorbed neon. This suggests that most of the Ne atoms will be adsorbed at the edges near the contact where the graphene-gold corner provides more advantageous adsorption conditions. The accumulation of neon to the edge can enhance the rigidity of the boundary condition, thereby increasing the frequency. However, it is also likely that the adsorbed neon will cause local strain, which enhances the frequency in spite of the increased mass. Similar behavior has been observed with $^3$He atoms on a carbon nanotube \cite{Igor2020}. 

The increase in strain is also corroborated in the shift of the minimum of the re-entrant looking $f_m(V_g)$ curve \cite{Song2011} under the influence of neon. The re-entrant W-shape indicates a crossover from the built-in-tension-dominated behavior to gate-induced-tension dependence, the position of which moves to larger $V_g$ with enhanced initial tension.

Strain variation on the atomic scale due to adsorbed atoms will lead to local pseudomagnetic fields as well as changes in the scalar potential which can strongly enhance the generation of $1/f^{\gamma}$ noise by Ne. We recently investigated magnetoresistance of the very same suspended Corbino disk \cite{Masahiro2020}. Those experiments indicated that Coulomb scatterers at small carrier densities close to the Dirac point become better screened in the presence of a magnetic field, which led to a growing shift of the Dirac point with increasing magnetic field. Even stronger screening of charged impurities, together with a Dirac point shift, is observed in the presence of Ne adsorbants. In fact, the mobility $\mu_0$ determined from magnetoresistance $\Delta R \propto (\mu_0 B)^2$ increased by 40\% in the presence of Ne atoms, which is in line with the increase in $\mu_{FE}$ found from conductance. The strain induced by Ne atoms may increase the carrier density in graphene, which then improves screening of Coulomb impurities \cite{Hwang2007} in particular near the Dirac point and the mobility grows. Alternatively, strain-induced pseudomagnetic fields localize charge in such a way that residual impurities become screened.
Thus, fluctuations in the strain-induced screening and pseudomagnetic fields are the central factors that make the influence of Ne  so pronounced in the low-frequency noise in our experiments. Also contact regions are affected by similar effects as discussed in Sects. I and V in the SI \cite{SI}.

\begin{figure}[t!]
	\centering
	\includegraphics[width=.45\linewidth]{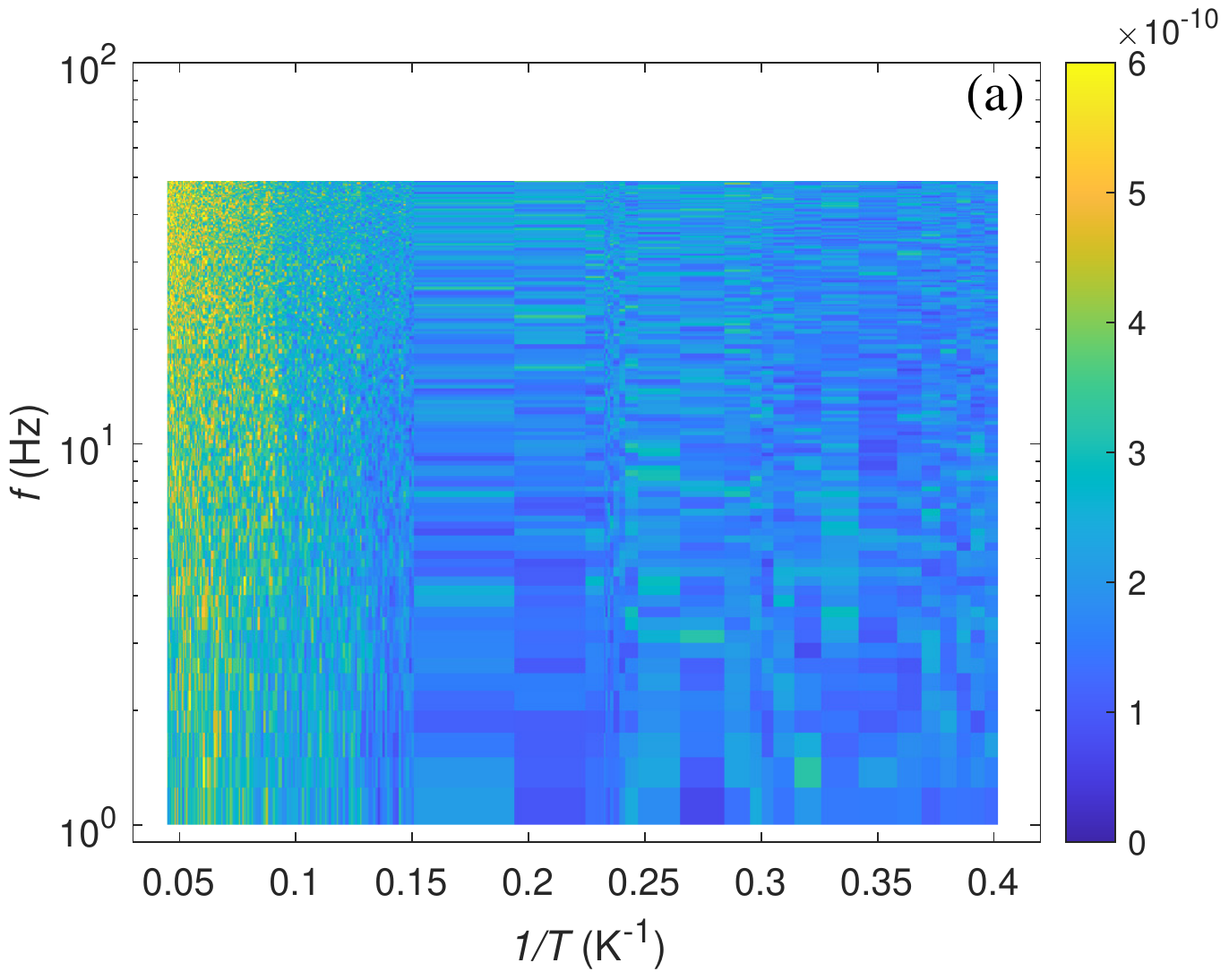}
	\includegraphics[width=.45\linewidth]{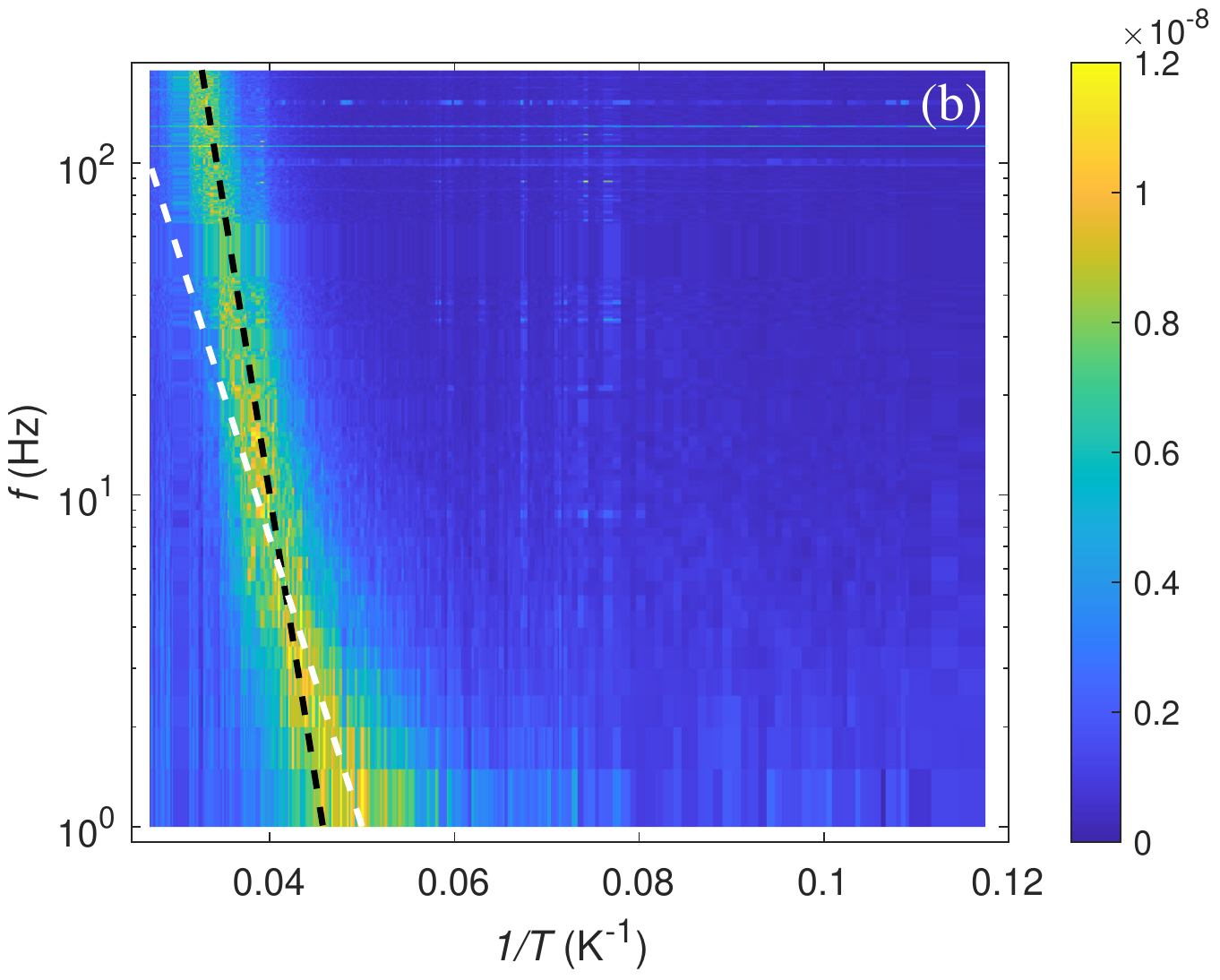}
	\caption{ a) Low frequency noise in terms of dimensionless product $f\times S_I/I^2$ for clean graphene on temperature $T$ vs. frequency $\log f$ plane; the scale is $\sim20\times$ lower compared to measurements with neon adsorbants. 
	b) Scaled noise $f\times S_I/I^2$  with neon atoms added to the sample depicted down to $T=8.5$ K in order to enhance the visibility of features at high $T$. The dashed lines illustrate the cut-off frequency $f_c$ (cf. Eq. (\ref{adsorption})) due to thermal activation, calculated for $f_c=f_0 \exp{(-E_a/k_B T)}$ with the adsorption energy of $E_a/k_B=410$ K (the steeper one) and with the surface trap potential $E_b/k_B=200$ K (the low temperature fit); SI Fig. 2 of SI \cite{SI} displays a schematic of the surface  potential with trapping states at the wall. The data were measured at the hole density of $|n| \sim 1.0 \times 10^{14}$\,m$^{-2}$. For the amount of Ne, see text. \label{fig:noise2d}}
\end{figure}

Fig. \ref{fig:noise2d}a displays low-frequency current noise measured on clean, suspended graphene. The data are displayed as $f \times S_I/I^2$ so that pure $S_I/I^2 \propto 1/f$ form becomes a constant in this plot. A Lorentzian fluctuator spectrum $\propto 1/(f^2+f_c^2)$, on the other hand, would display a peak at the corner frequency $f=f_c$. The data in Fig. \ref{fig:noise2d}a display an almost constant value at each temperature which means that the behavior is close to $1/f$ over the measured range which covers frequencies from 1 to 100 Hz while temperature is varied across $T=4-23$ K. 

The low frequency noise is substantially stronger with adsorbed neon as seen in Fig. \ref{fig:noise2d}b.  The data plotted as $f\times S_I/I^2$ display a maximum characteristic to a Lorentzian spectrum, in which the corner frequency $f_c$ moves exponentially with inverse temperature. Exponential behavior $\exp[-E_a/k_B T]$ is expected for thermally activated process following the Arrhenius law; here $k_B$ is the Boltzmann constant. Inspection of Fig. \ref{fig:noise2d}b indicates two activation processes with slightly different activation energies. The fitted lines yield $E_{a1}/k_B=410$ K and $E_{a2}/k_B=200$ K. We identify $E_{a1} = E_a$ as the adsorption energy of neon onto graphene, while the latter $E_{a2} = E_b$ is identified as describing trap states at the boundary (see Sect II of SI \cite{SI}). These trap states can act as expediters of trapping/detrapping behavior, potentially providing a similar resistance (current) noise mechanism as adsorption/desorption phenomena. Furthermore, the atomic graphene lattice potential has corrugations with saddle-point-like barriers separating nearby graphene hexagons. The ensuing diffusion barrier height for neon atoms amounts to approximately $E_d =32$ K based on the values for graphite \cite{Carlos1980,Cole1980}. Consequently, the observed current noise due to adsorbant dynamics on graphene is governed partly by surface diffusion with intermediate trapping and partly by adsorption/desorption, the relative weight of them depending on the temperature \cite{Vig1999,Yang2011}. 

Initially at low temperatures, we have thermally activated diffusion along the substrate which  becomes gradually influenced by desorption from the surface with increasing $T$. As discussed in the SI \cite{SI}, we think that adsorption from the gas phase to the graphene membrane is limited due to lack of sticking sites except at the electrodes. 
The trap states at the boundary feed atoms back to the graphene surface at a rate governed by the relevant Arrhenius law, namely $\propto \exp[-E_b/k_B T]$, where $E_b$ describes the depth of the trapping potential with respect to the potential surface in graphene (for a schematic picture, see SI \cite{SI}).

To describe the rate of change $\dot{N}$ in the number of the adsorbed Ne atoms on the graphene membrane, we employ a model with $N$ mobile adsorbed atoms and a supply of $N_b$ atoms captured by the additional trapping potential at the electrodes. Thus, our model has two coupled rate equations, one for $N$ and one for $N_b$ as outlined in the SI  \cite{SI}. The rates of exchange of atoms between the gas phase, the graphene membrane, and the surface trapping yield $N$ and its fluctuation rate that governs the low frequency scattering noise in the electronic transport. At high temperatures $T > 25$ K, desorption of atoms is faster than their diffusion and direct adsorption/desorption processes govern the current noise. At low temperatures, on the other hand, exchange of atoms with gas phase becomes irrelevant, and the dynamics of atoms is governed by release from the electrodes and ensuing diffusion on the Corbino disk. 

Adsorption/desorption processes lead to Lorentzian noise spectrum given by
\be \label{adsorption}
{S_I} = g N\frac{f_c}{f^2 + f_c^2},
\ee
where $g$ reflects the strength of individual scatterers, $N$ describes number of particles involved in the process, and $f_c$ is the frequency of desorption and adsorption processes which are equal at equilibrium. For thermal activation we may write $f_c=f_0 \exp{(-E_a/k_B T)}$ where $f_0$ is the attempt frequency and $E_a$ is the binding energy of Ne atoms on the graphene substrate. By fitting Eq. (\ref{adsorption}) to the data in Fig. \ref{fig:noise2d}b, we obtain $E_a/k_B = 410$ K with $f_0$ in the range of $ 10^8$ s$^{-1}$, which is rather low for an attempt frequency. However, low values for $f_0$ have been obtained for surface diffusion of noble gases on metallic surfaces \cite{Barth2000}. Compared with neon adsorption energy on graphite $E_a/k_B=350$ K \cite{Antoniou1976}, our value is reasonable, taking into account the possible increase in interaction energy due to local deformation in a single layer substrate. We emphasize that we always see a single Lorentzian line in the adsorption/desorption regime, never a collection of two level systems as seen for example in high-Ohmic graphene tunneling devices \cite{Puczkarski2018}.

With lowering temperature well below 25 K, the desorption rate of atoms becomes very small. Using the exponential activation fit to Fig. \ref{fig:noise2d}b, the desorption rate ${f_{ds}} = {f_0}\exp \left(-\frac{E_a}{k_B T} \right)$ becomes $\sim 0.1$ mHz at $T=15$ K (with $f_0=10^8$ s$^{-1}$). Diffusion time of Ne atoms across the Corbino ring varies between $\tau_d=250 \dots 2$ ms at temperatures $T= 4 \dots 10$ K (see Sect. 2 of SI  \cite{SI}). Consequently, the probability of desorption from surface during diffusion ${P_{ds}} \cong {f_0}\exp \left(-\frac{E_a}{k_B T} \right) \times {\tau _d} \ll 1$, and Ne atoms may diffuse across the Corbino disk without being desorbed during their flight time at $T < 10$ K. Therefore, diffusion of Ne atoms will govern fluctuations in the absorbent number and configuration patterns, which in turn, govern the low-frequency resistance noise in our system. The same diffusion processes on graphene govern also fluctuations in the number of atoms at the edge. This will lead to fluctuations in the contact noise, the separation of which is discussed in Sect. V in the SI \cite{SI}. 

The diffusion process leads to a random walk type of noise where the characteristic frequency is given by the inverse of a typical random walk time \cite{Voss1976a,Yakimov1980}, \emph{i.e.} the inverse of diffusion time across the sample $f_c=(2\pi\tau_d)^{-1}$ and $N$ in Eq. (\ref{adsorption}) reflects now the average fluctuating number of atoms diffusing along the graphene sample. Hence, at $T=4-10$ K, surface diffusion provides low-frequency noise in the range of investigated frequencies $1-100$ Hz. However, the diffusion of individual, non-interacting particles will lead to a noise spectrum of $1/f^{1.5}$ form \cite{Yakimov1980,Scofield1985,Zimmerman1988,Kar2002}. Only if there are additional correlations, for example generated by clustering/declustering of neon atoms via thermally driven surface diffusion, the noise power spectral density may approach the $1/f$ spectrum. Basically, multipartite clustering dynamics leads to long-term memory effects, which modify the random walk nature of regular low-frequency diffusion noise. 

\begin{figure}[t!]
	\centering
	\includegraphics[width=.49\linewidth]{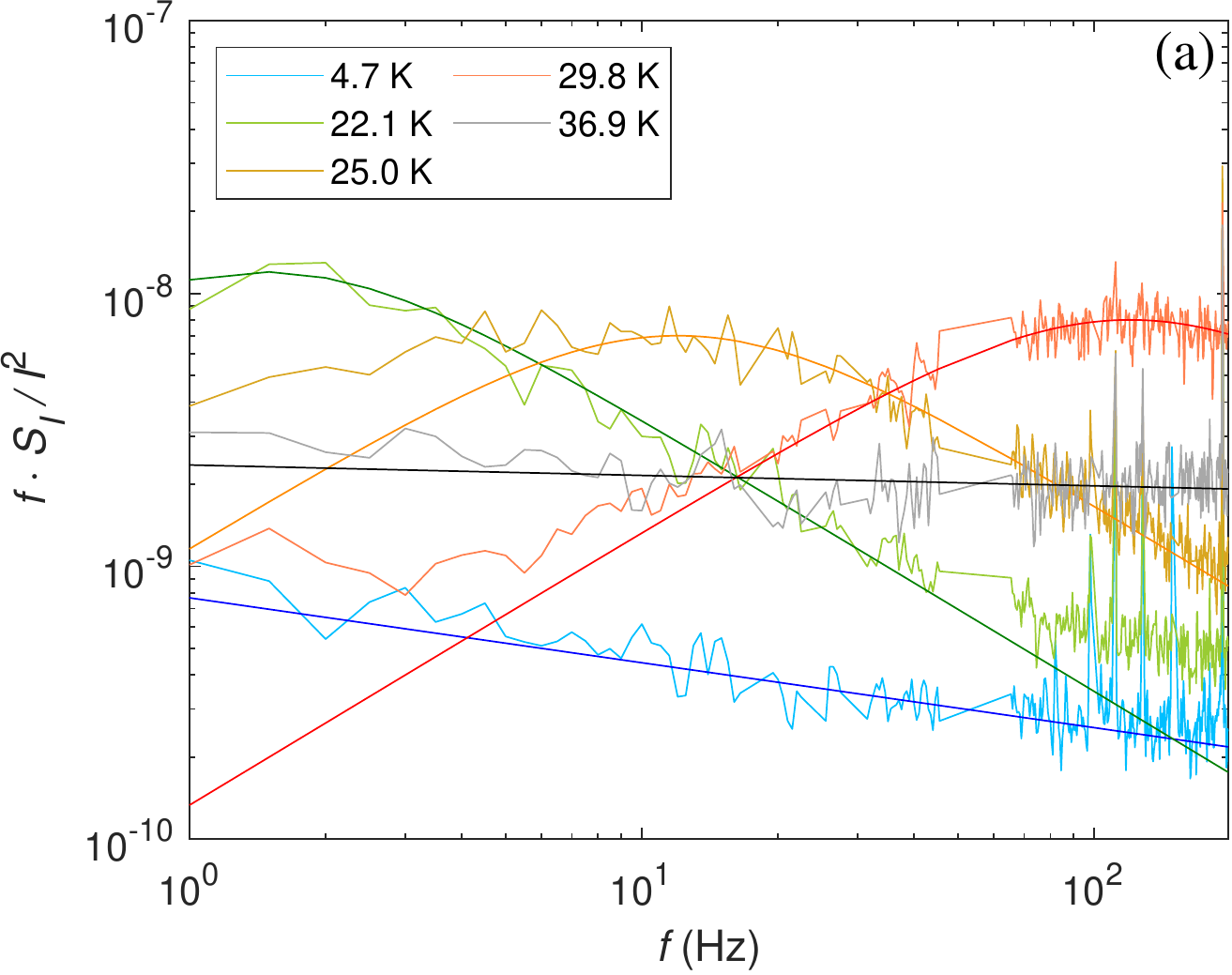}
	\includegraphics[width=.49\linewidth]{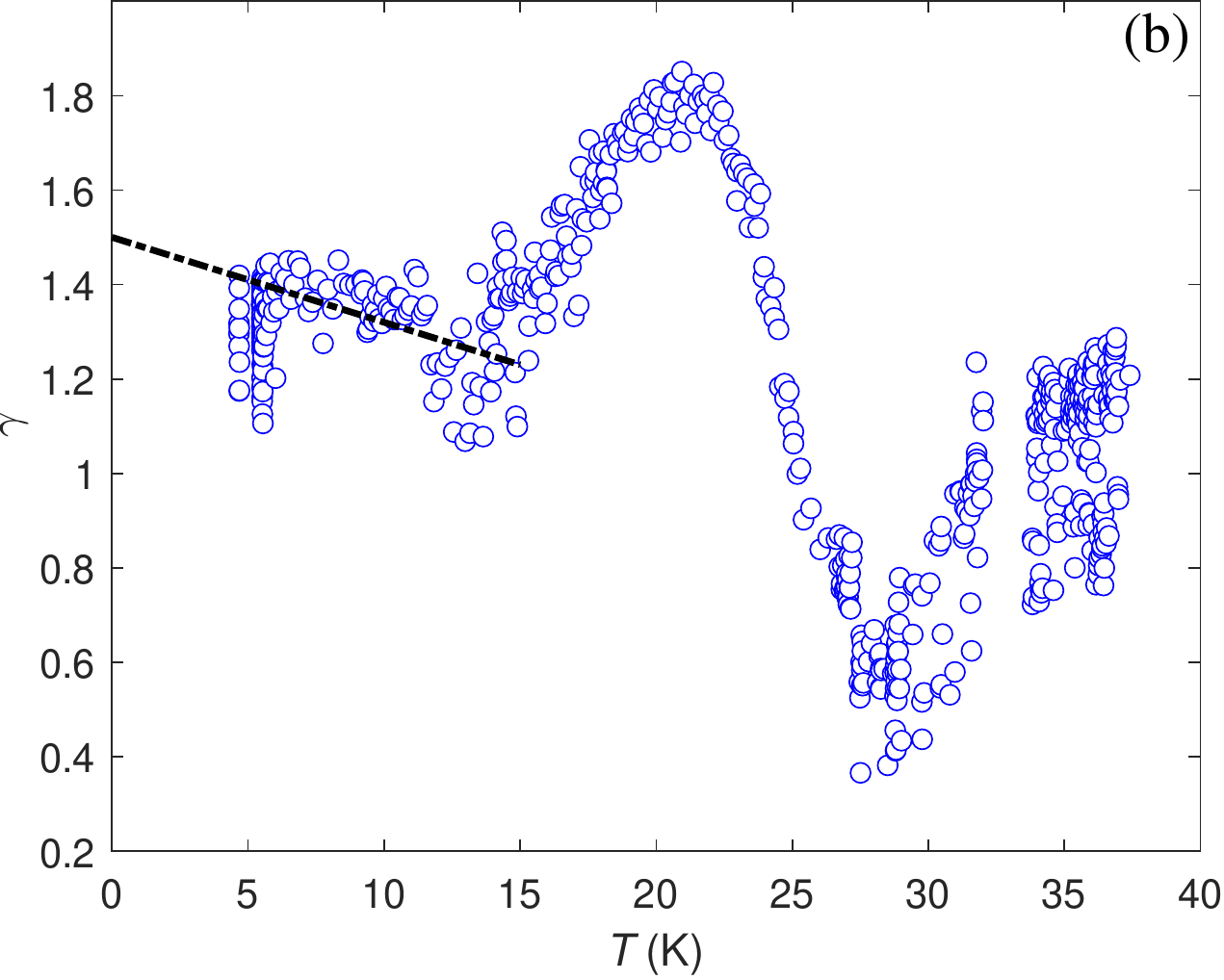}
	\caption{a) Current noise spectral density multiplied by frequency $f \times S_I/I^2$ measured at temperatures $T= 4.7$ K, 22.1 K, 25.0 K, 29.8 K, and 36.9 K. The three intermediate temperature traces are listed from bottom to top with regards to the high frequency end (green, orange, red), with the overlaid traces obtained from Eq. (\ref{adsorption}) using $f_c= 1.5$, 12, and 120 Hz, respectively. 
	The data at $T=4.7$ K (blue) does not include a Lorentzian part and it is described by a spectrum $1/f^{\gamma}$ with $\gamma =  1.24$, while $\gamma=1.04$ for the data at 36.9 K (gray/black).
	b) Exponent $\gamma$ obtained from an overall $S_I \propto 1/f^{\gamma}$ fit to the data in Fig. \ref{fig:noise2d}b. The movement of $f_c$ across the studied frequency range as a function of $T$ is visible here as a change in $\gamma$ from 1.8 to 0.5, equalling nearly the  change from 2 to 0 expected from Eq.  (\ref{adsorption}). The dashed line sketches clustering-induced decrease of $\gamma$ vs. $T$.  \label{NoiseFIT}}
\end{figure}

In order to detail the noise in the desorption regime, Fig. \ref{NoiseFIT}a displays a few current noise spectra as $f \times S_I/I^2$ measured at different temperatures between $4-37$ K. With increasing temperature, the peak in $f \times S_I/I^2$ due to the cut-off frequency $f_c$ shifts towards higher frequencies. The overlaid traces at temperatures $T=22.1$ K, $T=25.0$ K, $T=29.8$ K are fits calculated using Eq. (\ref{adsorption}) with $f_c= 1.5$, 12, and 120 Hz, respectively. These values correspond  to $f_c=f_0 \exp{(-E_a/k_B T)}$ with the attempt frequency $f_0=1.4 \pm 0.25 \times 10^8$ s$^{-1}$. In general, the fits in Fig. \ref{NoiseFIT}a are  remarkably accurate, although at intermediate temperatures the low frequency part deviates from a full Lorentzian, and a noise contribution with spectral density of $1/f^{\gamma}$ form starts to push out from the data trace (see the flattening parts at low frequency in the traces at $T= 25.0$ K and 29.8 K in Fig. \ref{NoiseFIT}a).
Fig. \ref{NoiseFIT}a displays also one spectrum at high temperatures, at $T=36.9$ K, at which the adsorption/desorption phenomena are so fast that we are left only with $1/f$ noise.

A distinct feature in the data of Fig. \ref{NoiseFIT}a is the absence of temperature dependence in the maximum amplitude of the $f \times S_I/I^2$ traces at temperatures $T=22.1$ K, $T=25.0$ K, $T=29.8$ K. This arises from the specific behavior of the sticking sites at the boundaries, \emph{i.e.} due to the fact that the feed from the $N_a$ trapped atoms is able to compensate the desorption rate of $\sim N$ so that $N=\mathrm{const}.$ 
%(see SI  \cite{SI}). 
If $N$ would be governed by uniform graphene surface adsorption/desorption processes with a moderate sticking probability, a strong decrease of the maximum of $f \times S_I/I^2$ would be observed with lowering $T$.  

Besides the Lorentzian based fits of Fig. \ref{NoiseFIT}a, we made power law fits to the noise  using an arbitrary exponent: $S_I \propto 1/f^{\gamma}$. Results of these unconditional fits to the data are illustrated in Fig. \ref{NoiseFIT}b. The movement of $f_c$ across the studied frequency range as a function of $T$ is visible here as a non-monotonic change in $\gamma$.  A change in the exponent from 2 to 0 is expected on the basis of Eq.  (\ref{adsorption}), but the data display a smaller swing: from 1.8 to 0.5. Adopting a commonly used criterion $0.5 < \gamma < 1.5$ for $1/f$ noise, we may conclude that the presence of Ne is able to destroy the character of the noise when the sloped noise part $\propto 1/f^2$ is dominating. At $T=4-10$ K, the noise spectrum is described by a single exponent $\gamma = 1.2 - 1.4$. One spectrum with $\gamma = 1.24$ measured at $T=4.7$ K is displayed for reference in Fig. \ref{NoiseFIT}a.

Noise induced by thermal diffusion has been demonstrated to lead to complex frequency dependence of noise\cite{Voss1976}. In our hopping transport, the complexity of neon diffusion depends on Ne-Ne interactions and the boundary conditions. If one calculates numerically the life time distribution of diffusing particles in a Corbino disk geometry with fully absorbing boundary conditions (see Sect. III of SI  \cite{SI}), one obtains a noise spectrum of the form $1/f^{1.5}$, similar to the one dimensional case \cite{Yakimov1980}.  On the other hand, if we assume a probability of reflection of particles from the boundary (the relevant sticking site occupied), then we obtain $\gamma > 1.5$. Thus, without additional assumptions on time dependent correlations among neon atoms, for example changes in scattering via clustering/declustering, the diffusion model is not sufficient for explaining the observed exponent $\gamma=1.2-1.4$.

The central question of low-frequency noise due to diffusing adsorbants is the nature and strength of their mutual interactions. As discussed in Sect. I of the SI  \cite{SI}, it is known that Ne atoms have an attractive interaction on the order of $\epsilon/k_B \simeq 40$ K which tends to stabilize $ \sqrt{7} \times \sqrt{7}$ commensurate structures on graphite. Thus, clusters of Ne atoms on graphene may form and they modify random diffusion by their interaction energy and by hopping restrictions imposed on the atoms sitting within the cluster. We have performed kinetic Monte-Carlo simulations (see Sect. IV in SI \cite{SI}) to investigate these effects and have found that the significance of clusters depends on temperature. At higher temperatures more atoms are diffusing and the interactions and clusters become more important. According to these MC simulations, the noise is first close to $1/f^{1.6}$ at low $T$, at which mostly single atoms are diffusing, but it becomes closer to $1/f^{1.2}$ when temperature is increased by a factor of approximately two. These simulation results agree quite well with our data in Fig. \ref{NoiseFIT}b, in which a decrease of $\gamma$ from 1.4 to 1.2 is observed when temperature is varied between $T=4-10$ K. This agreement strongly supports the importance of clustering of the adsorbants for generation of $1/f$ type of noise. 

In summary, we have investigated low-frequency noise in suspended graphene with and without adsorbed neon, in particular, in Corbino geometry in which there are no free edges to interact with adsorbed gas atoms. We find ultra-low $1/f$ noise amounting to $f \cdot S_I/I^2 = (2 \pm 0.5) \times 10^{-10}$ for clean graphene at intermediate charge densities $n = \pm 7 \cdot 10^{11}$ cm$^{-2}$ at 4 K; the noise was enhanced by a factor three when adding neon on the sample at these carrier densities. The added neon amount increased the mechanical resonant frequency by 2 \textperthousand, indicating additional Ne-induced strain in the membrane.

At $T>>10$ K, desorption of Ne atoms led to fluctuations in the number of mobile Ne atoms on the graphene surface which caused a temperature-dependent characteristic fluctuation frequency $f_c=f_0 \exp{(-E_a/k_B T)}$ that corresponds to adsorption energy $E_a/k_B=410$ K. At $T < 20$ K, the noise is governed by surface diffusion of Ne atoms and models based on dynamical clustering of mobile impurities were tested at $T=4-10$ K. Our work clearly demonstrates that substantial amount of low-frequency noise is created by diffusing impurities on a 2D sample. 
The observed noise spectra around $4-10$ K display a power law behavior $1/f^{\gamma}$ with $\gamma = 1.2 - 1.4$.
Fundamental agreement with our Monte-Carlo simulations supports the conclusion that weakening of the frequency exponent from the single-particle diffusion noise with $\gamma=1.5$ towards $\gamma = 1$ is due to the variation in scattering cross section caused by clustering/declustering of the mobile neon impurities. Our results carry direct relevance for ultra clean graphene technologies \cite{Zhang2020} and they provide strong evidence that diffusing defects and their relative grouping/regrouping play a role in various systems displaying $1/f$ type of noise.

\begin{acknowledgement}
 Discussions and correspondence with Vanessa Gall, Igor Gornyi, Adrian del Maestro, Liu Ying, Manohar Kumar, Elisabetta Paladino, Sergey Rumyantsev, and Igor Todoshchenko are gratefully acknowledged. This work was supported by the Academy of Finland projects 314448 (BOLOSE), 310086 (LTnoise) and 312295 (CoE, Quantum Technology Finland) as well as by ERC (grant no. 670743). This research project utilized the Aalto University OtaNano/LTL infrastructure which is part of European Microkelvin Platform (funded by European Union’s Horizon 2020 Research and Innovation Programme Grant No. 824109). A.L.\ is grateful to V{\"a}is{\"a}l{\"a} foundation of the Finnish Academy of Science and Letters for scholarship.
\end{acknowledgement}

\begin{suppinfo}
The following files are available free of charge.
\begin{itemize}
  \item Supplementary Information.pdf: supplementary information
  \item mc140\_movie\_small\_dt.mp4: Monte Carlo simulation at $k_BT=1.2$
  \item mc147\_movie\_small\_dt.mp4: Monte Carlo simulation at $k_BT=2$
\end{itemize}

\end{suppinfo}

\newpage
\begin{center}
\section*{{\small SUPPLEMENTARY MATERIALS}} \label{SI}

\textbf{Electrical low-frequency $1/f^{\gamma}$ noise due to surface diffusion of scatterers on an ultra low noise graphene platform}\\
\vspace{1cm}
%\author{authors}
Masahiro Kamada$^1$, Antti Laitinen$^1$, Weijun Zeng$^1$, Marco Will$^1$, Jayanta Sarkar$^1$, Kirsi Tappura$^2$, Heikki Sepp\"a$^2$, Pertti Hakonen$^1$\\
\vspace{0.5cm}
\small{$^1$Low Temperature Laboratory, Department of Applied Physics, Aalto University\\ School of Science, P.O. Box 15100, 00076 Aalto, Finland}\\
\small{$^2$Quantum systems, VTT Technical Research Centre of Finland Ltd.,\\ P.O. Box 1000, 02044 VTT, Finland}
\end{center}

    \section*{I. Neon on graphite}
   Atomic neon films on graphite provide a good starting point for understanding our submonolayer Ne films on graphene. An excellent overview of neon films on graphite is provided in Ref. \citenum{Bruch2007}. In general, quantum effects are a way smaller for neon atoms than for helium or hydrogen but they are
still signiﬁcant. For example, diffusion of Ne atoms is governed by quantum tunnelling at subkelvin temperatures but, in our work at $T>4$ K, thermally activated diffusion still dominates.

According to Ref. \citenum{Gatica2009}, Ne-Ne interaction has a binding energy of $\epsilon/k_B=42$ K, while the binding energy to graphite $E_a/k_B=380$ K, pretty close to the experimentally measured value of Ref. \citenum{Antoniou1976}. Casimir-like adatom-adatom interaction via graphene electrons is also possible \cite{Shytov2009}. For commensurate solid structures, neon prefers a registered phase with $\sqrt{7} \times \sqrt{7}$ lattice and a four atom basis  \cite{Huff1976,Bruch2007}. This means that interatomic interaction energy is minimized by local order, and that self-bound ordered regions, clusters, become possible, even though they would just provide a temporary structure. Parallel to this, the binding energy between neon atoms can be on the order of the diffusion barrier $E_d$, which means a significant energy scale considering the surface dynamics of atoms. These considerations provide justification for the parameter values employed in our kinetic Monte Carlo simulations targeting the clustering dynamics of adsorbed neon atoms (see Sect. IV below). 

Our adsorbed neon films were prepared around 20 K with pressure $p \sim 10^{-4}$ mbar. On the basis of the empirical phase diagram for thin Ne films on graphite \cite{Calisti1982,Bruch2007}, we conclude that no registered phases will be generated, and the behavior of the adsorbed atoms at $T=10-37$ K will display gas or fluid like behavior. At temperatures $T < 10$ K, coexistence of solid phase and vapor becomes possible \cite{Calisti1982}. An immobile solid phase will be irrelevant for the generation of low frequency noise. Fluctuations of the shape of the solid, however, will contribute to the noise in a similar fashion as fluctuations in the clustered atom regions seen in our simulations.

\begin{figure}[b!]
	\centering
	\includegraphics[width=.48\linewidth]{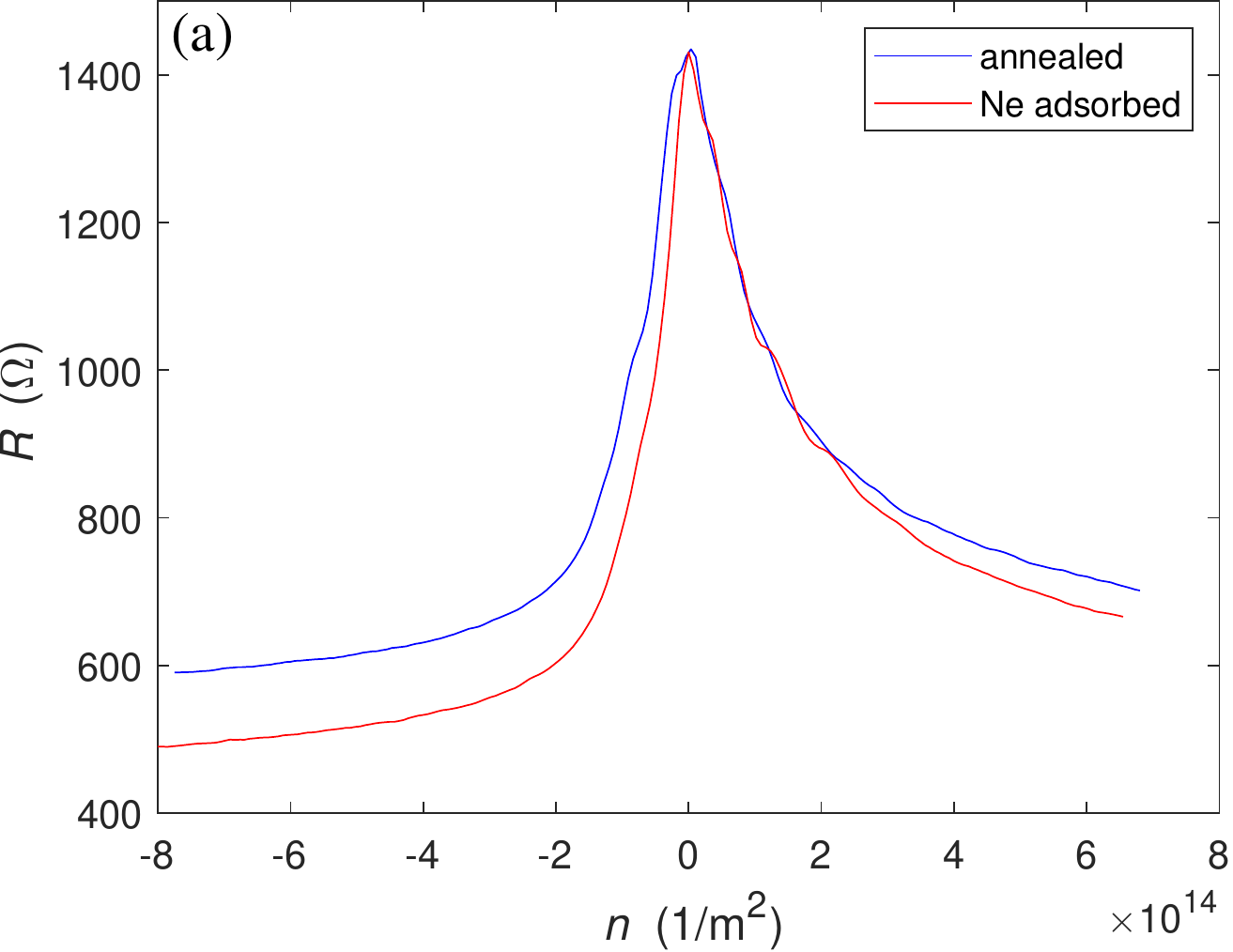}
	\includegraphics[width=.48\linewidth]{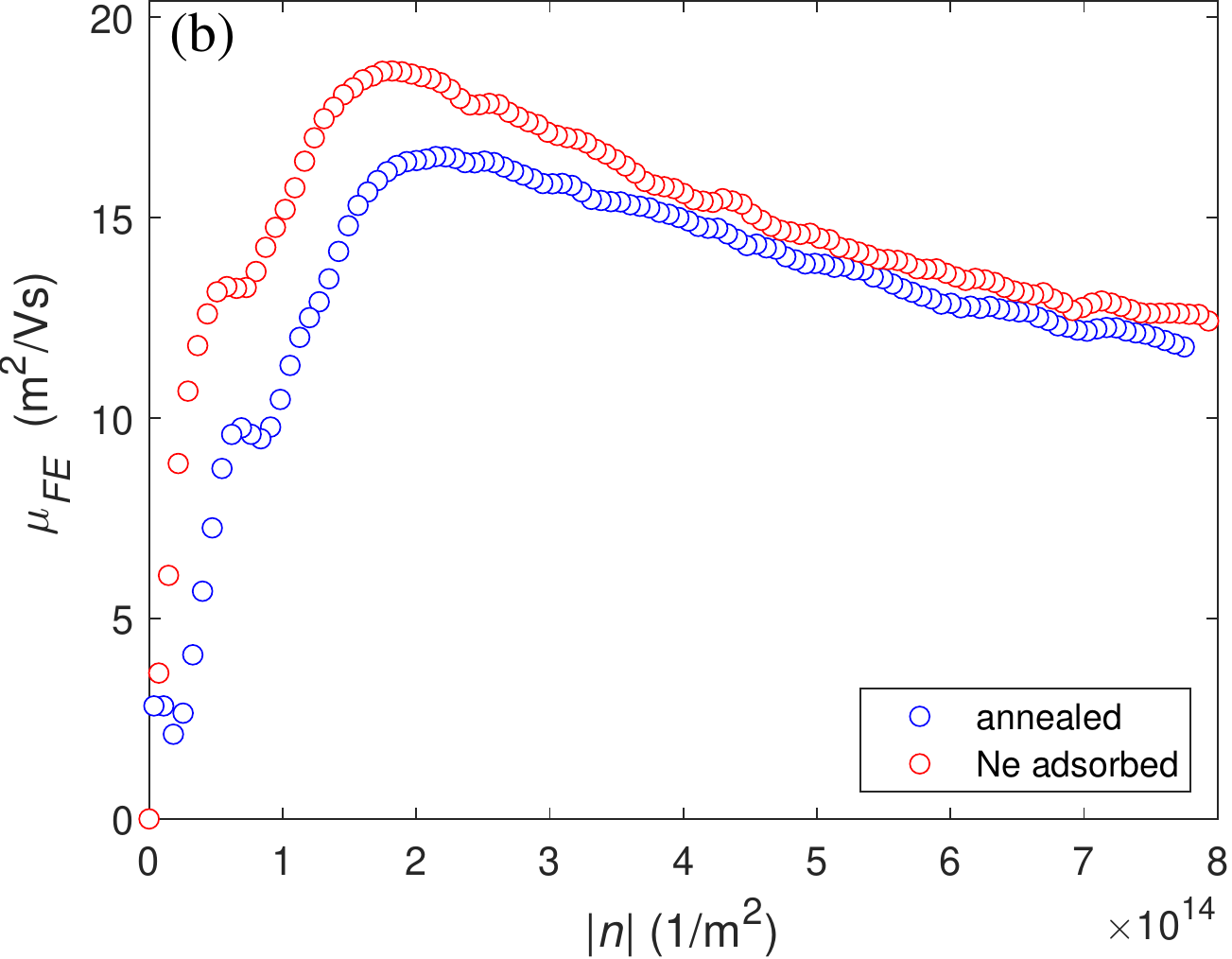}
	
	\caption{(Left) Zero bias resistance $R$ of the Corbino disk as a function of charge carrier density $n$ measured in a state with adsorbed neon (lower trace at left) and in a subsequent annealed, clean state (upper trace at left). (Right) Field effect mobility $\mu_{FE}$ for holes obtained from the resistance data on the left at $n<0$.
	  \label{SI:fig:RvsVg}}
\end{figure}

The addition of neon changed the dependence of graphene sample resistance, $R$, on the gate-induced carrier density $n$ as illustrated in Fig. \ref{SI:fig:RvsVg}a. Clearly, $R(n)$ is more peaked near the Dirac point  while the saturation conductance (inverse resistance) value becomes larger with Ne adsorption; this saturation value at $|V_g| \gg V_g^D$ is taken to correspond to the effective contact resistance $R_c$. Consequently, the data in Fig. \ref{SI:fig:RvsVg}a yields $R_c = 500$\,$\Omega$ and $405$\,$\Omega$ for clean and Ne-adsorbed states, respectively. With addition of Ne, the Dirac point of the sample also moved upwards nearly by 0.8 V in $V_g$, which has been taken into account when calculating the charge density.

Fig. \ref{SI:fig:RvsVg}b displays the field effect mobility calculated from $\mu_{FE}(n)=\left[\sigma(n) -\sigma_0(n_0)\right]/ne$ in which $\sigma(n)$ and $\sigma_0(n_0)$ denote the conductivity of the sample at carrier density $n$ and at residual carrier density $n_0$ at the Dirac point.
We display $\mu_{FE}(n)$ only for positive carriers because then $R$ is not influenced by interfaces in the pnp doping structure at $V_g > V_g^D$. In the analysis, we determine separately the contact resistance for clean and Ne-adsorbed states and  subtract $R_c$ off from $R$ before calculating the conductivity. Data in Fig. \ref{SI:fig:RvsVg}b indicate that addition of neon  increases $\mu_{FE}$ by 30\% at carrier densities $n = 1 \dots 2 \times 10^{14}$\,m$^{-2}$ which is the range where most of our noise data have been measured. This increase in mobility is quite close to the quoted 40\% increase in $\mu_0$ found from the magnetoresistance.

The asymptotic resistance value at large negative gate voltages in Fig. \ref{SI:fig:RvsVg}a is reduced by $\sim 95$ $\Omega$ by the addition of Ne. This indicates that resistance for hole transport near the gold contact is influenced by the adsorption of Ne and the hole contact resistance $R_c^h$ (resistance not influenced by $V_g$) is lowered. Reduction in $R_c^h$ can be assigned to Ne-induced strain that causes a change in the scalar and vector potentials in graphene in the neon-covered region. The strain-induced scalar potential modifies charge density near the contact and $R_c^h$ becomes lowered as seen in Fig. \ref{SI:fig:RvsVg}a. For electronic transport at $V_g >> V_g^D$, the modification in the electronic contact resistance $R_c^e$ is nearly zero.

%The addition of neon changed the graphene sample resistance as illustrated in Fig. \ref{SI:fig:RvsVg}. In particular, the resistance  increased by approximately 10\% while the Dirac point of the sample moved upwards nearly by 0.8 V in $V_g$. The calculated field effect mobility $\mu_{FE} \propto \mathrm{d}(1/R)/\mathrm{d}V_g$ increased by $\sim 50$\% by the addition of neon. This increase is quite close to the quoted 40\% increase in $\mu_0$ found from the magnetoresistance. \ph{We also observe that the asymptotic resistance value at large negative gate voltages is reduced by $\sim 60$ $\Omega$. This indicates that resistance for hole transport near the gold contact is influenced by the adsorption of Ne and the hole contact resistance $R_c^h$ (resistance not influenced by $V_g$) is lowered. Reduction in $R_c^h$ can be assigned to Ne-induced strain that causes a change in the scalar and vector potentials in graphene in the neon-covered region. The strain-induced scalar potential modifies charge density near the contact and $R_c^h$ becomes lowered as seen in Fig. \ref{SI:fig:RvsVg}. For electronic transport at $V_g >> V_g^D$, the modification in the electronic contact resistance $R_c^e$ is nearly zero. }

\section*{II. Thermal activation and potentials}

\begin{figure}[b!]
	\centering
	\includegraphics[width=.60\linewidth]{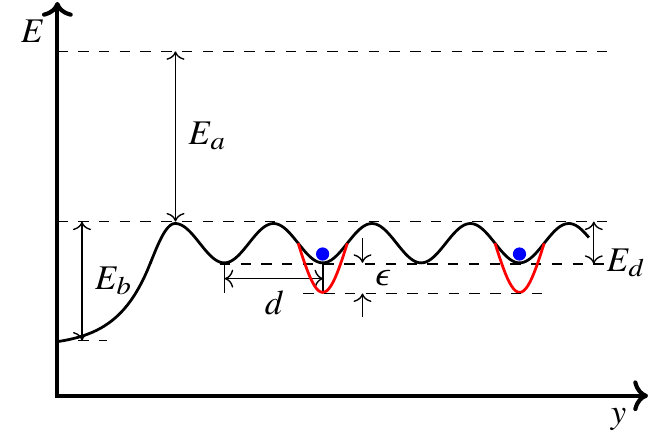}
	
	\caption{ Schematic potential for neon atoms on graphene near the gold electrode. Horizontal axis $y$ denotes the distance from the gold wall. The energy scales, $E_a$, $E_b$, and $E_d$ are explained in the text. The scale $d = 2.46$ \AA $ $ marks the distance between the nearest hexagons of the graphene lattice. The two Ne atoms on the graphene lattice interact with energy $\epsilon$ which is the source of clustering, opposed by declustering due to thermal agitation.
	  \label{potential}}
\end{figure}

Our basic model for atomic Ne fluxes and   trapping of atoms is illustrated in Fig. \ref{potential}.
The joint dynamics of gaseous neon and adsorbed atoms is governed by three different energy scales. The largest is the Ne-graphite adsorption energy $E_a/k_B$ which is on the order 350 K \cite{Antoniou1976}. For the atoms at the electrodes of the Corbino geometry there is an additional Ne-boundary adsorption energy $E_b/k_B$ which amounts approximately to a few hundred Kelvin which is a typical noble gas - metal adsorption energy \cite{Barth2000}. The smallest scale is related to the corrugation of the graphite potential $E_d$ that provides the diffusion barrier of the neon atoms moving along graphene. We employ $E_d/k_B = 32$ K which has been reported for graphite \cite{Carlos1980,Cole1980}. 
In our simple modeling, we neglect the Ne-Ne interaction, but our Monte Carlo simulations do take this energy into account (see Sect. IV).

When Ne atoms land on clean graphene, there are very few sticking sites for the atoms to get attached before they reach the boundary of the sample. Consequently, in our modeling we assume that neon atoms, becoming adsorbed from the gas phase, will become attached first to the wall potential marked by $E_b$ in Fig. \ref{potential}. We are interested in the time development of the number of mobile Ne atoms $N$ on the graphene surface. Since neon atoms diffusing on graphene are supplied by $N_b$ trapped atoms at the walls, we have two rate equations governing the dynamics:
\begin{eqnarray}
{{\dot N}_b} =  - \frac{{{N_b}}}{{{\tau _s}}} + \dot N_a^{(gas)} ,\\
{{\dot N}} =  - \frac{{{N}}}{{{\tau _{ds}}}} - \frac{{{N}}}{{{\tau _d}}} + \frac{N_b}{{{\tau _s}}} \label{rate},
\end{eqnarray}
where the upper equation describes the behavior at the wall and the latter one on graphene. $N_b/\tau_{s}$ denotes the rate at which atoms are released from the wall while $\dot N_a^{(gas)}$ gives the influx of atoms to the wall from the gas phase. The rate of change of Ne atoms on graphene ${{\dot N}}$ is given by desorption $-N/\tau_{ds}$, by diffusion and retrapping to walls $-N/\tau_{d}$, and by release from the walls at rate $N_b/\tau_{s}$, where $\tau_i$ with $i =\{d,s,ds\}$ denotes the characteristic time scale for diffusion across the sample and retrapping to the walls, life time at the wall, and life time on graphene before desorption, respectively. 

In the steady state, $\frac{{d{N_b}}}{{dt}}=0$ and $\frac{{d{N}}}{{dt}}=0$, which yields 
\begin{eqnarray}
N_b= \dot N_a^{(gas)} \tau_s ,\\
N =  \frac{{{\tau _{ds}}{\tau _d}}}{{{\tau _{ds}} + {\tau _d}}}\frac{1}{{{\tau _s}}}N_b .
\end{eqnarray}

Using the definitions in Fig. \ref{potential}, the time scales can be written as 
\begin{eqnarray}
\tau_s = \tau_s^{(0)}\exp \left( {{{{E_b}} \mathord{\left/
 {\vphantom {{{E_b}} {{k_B}T}}} \right.
 \kern-\nulldelimiterspace} {{k_B}T}}} \right),\\
{\tau _{ds}} = \tau _{ds}^{(0)}\exp \left( {{{{E_a}} \mathord{\left/
 {\vphantom {{{E_a}} {{k_B}T}}} \right.
 \kern-\nulldelimiterspace} {{k_B}T}}} \right),\\
{\tau _d} = \tau _d^{(0)}\exp \left( {{{{E_d}} \mathord{\left/
 {\vphantom {{{E_d}} {{k_B}T}}} \right.
 \kern-\nulldelimiterspace} {{k_B}T}}} \right).
\end{eqnarray}
All time scales involve exponential activation type of behavior, but with quite distinct energy scales. The prefactors are related to inverse attempt frequencies $f_0^{-1}$; these frequencies  are expected to be in the range of $f_0^{th} \simeq 10^8 - 10^{12}$ s$^{-1}$, but they vary depending on the actual curvature of the underlying potential and the temperature of the environment. However, since effective attempt frequencies have been observed to be even smaller than $f_0^{th} $ with noble gases on metal surfaces \cite{Barth2000}, we regard $\tau_s^{(0)}$ and $\tau _{ds}^{(0)}$  as fit parameters.

Diffusion of Ne atoms along the graphene is also thermally activated \cite{Barth2000} and depends exponentially on temperature according to
\be
D = {D_0}\exp \left(  - \frac{E_d}{k_B T} \right),
\ee
where 
$D_0$ denotes the prefactor which is expected to be on the order of $D_0 \sim \lambda^2 f_0^h$ in terms of the hopping length $\lambda$ and the attempt frequency for hopping $f_0^h$. Using $f_0^h=3.3 \cdot 10^{11}$ s$^{-1}$ and $\lambda=2.46$ \AA, the hexagon-hexagon spacing, one obtains a value $D_0 \simeq 2.0 \times 10^{-8}$ m$^2$/s.  Thus, we may estimate for the diffusion coefficient  $D=6.7 \times 10^{-12}$ m$^2$/s ($D=8.1 \times 10^{-10}$ m$^2$/s) at 4 K (10 K) and the diffusion time becomes
${\tau _d} = \frac{{{L^2}}}{D} = \frac{{{L^2}}}{{{D_0}}}\exp \left( { + \frac{{{E_d}}}{{{k_B}T}}} \right)=250$ ms (2 ms).

At high temperatures $T > 25$ K, desorption of atoms is faster than their diffusion across the sample, and we have $\tau_{ds} \ll \tau_d$, while the opposite limit $\tau_{ds} \gg \tau_d$ is realized at low temperatures. In the former case, we have
\be
N = \frac{{{\tau _{ds}}{\tau _d}}}{{{\tau _{ds}} + {\tau _d}}}\frac{1}{{{\tau _s}}}{N_b} \simeq \frac{{{\tau _{ds}}}}{{{\tau _s}}}{N_b} = {\tau _{ds}}\dot N_a^{(gas)} \simeq \mathrm{const.}
 ,
\ee
because $\tau_{ds}$ and $\dot N_a^{(gas)}$ have similar exponential temperature dependence.
This is in accordance with the observed $T$-independence of the Lorentzian noise. 
Consequently, the sticking sites in our graphene device do not reside on the Corbino membrane, but at the boundary and they provide an effective, external ballast for the number of atoms of the graphene sheet as assumed in the analysis of the main text.

In the low-temperature limit with $\tau_{ds} \gg \tau_d$, all the atoms from the gas phase have been adsorbed to surfaces, and we expect that the trapping states at the boundary are basically fully occupied, \emph{i.e.} $N_b =N_S$, where $N_S$ denotes the saturation amount at the wall.
\be
N \simeq \frac{\tau _d} {\tau_s} N_S = \tau _d^{(0)}\exp \left( {{{{E_d}} \mathord{\left/
 {\vphantom {{{E_d}} {{k_B}T}}} \right.
 \kern-\nulldelimiterspace} {{k_B}T}}} \right)\frac{N_S}{{\tau _s^{(0)}}}\exp \left( {{{ - {E_b}} \mathord{\left/
 {\vphantom {{ - {E_b}} {{k_B}T}}} \right.
 \kern-\nulldelimiterspace} {{k_B}T}}} \right) =  \frac{{\tau _d^{(0)}}}{{\tau _s^{(0)}}}N_S\exp \left( {{{({E_d} - {E_b})} \mathord{\left/
 {\vphantom {{({E_d} - {E_b})} {{k_B}T}}} \right.
 \kern-\nulldelimiterspace} {{k_B}T}}} \right). \label{NlowT}
\ee
Since $E_d-E_b \sim -E_b$, there should be a strong $T$-dependence in $N$, and the number becomes reduced with lowering temperature. This will lead to more infrequent encounters between neon atoms and clustering of atoms becomes reduced at low $T$. This would then favor random walk type of noise which is detailed in Sect. III next. 

Finally, let us make a remark concerning the trapping potential $E_b$, which we have regarded as a constant. However, the potential at the boundary will have several energy levels for Ne atoms and the atoms on higher levels will experience a smaller trapping potential. 
Consequently, the effective trapping barrier $E_b^{eff}$ for release of Ne atoms at low $T$ may be smaller than $E_b=200$ K, which would lead to a reduced decrease of $N$ with lowering $T$ than obtained from Eq. \ref{NlowT}.

   \section*{III. Random walk on Corbino disk geometry}
\begin{figure}[H]
    \centering
    \includegraphics[width=\linewidth]{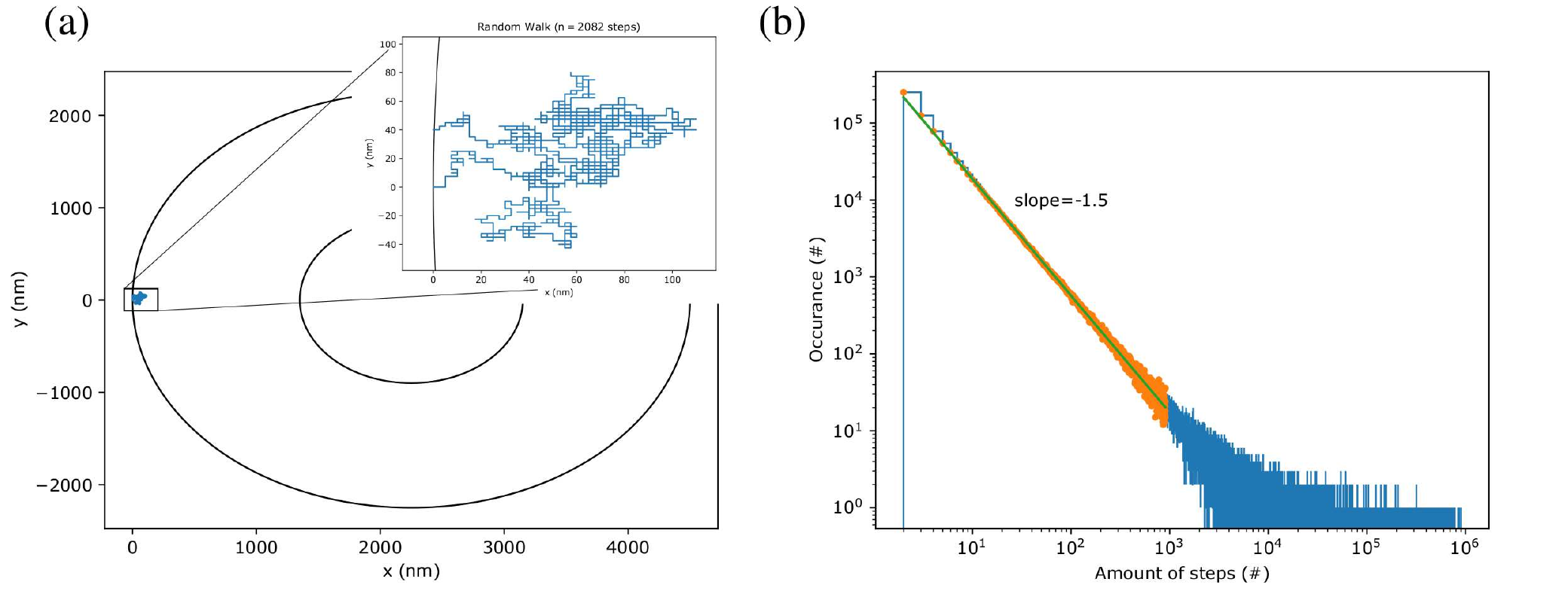}
    \caption{a) Geometry of the Corbino disk simulation and example of a random walk. b) Distribution of $1000000$ random walks (blue) on a $2.5\,$nm grid. Linear fit (green) to the first 1000 points (orange)  yields a slope of $-1.5$.}
    \label{Sup:fig:RandomWalk}
\end{figure}

Random walk simulations were performed with a Corbino disk geometry of two concentric circles of $1.8\,\upmu$m (inner) and $4.5\,\upmu$m (outer) forming the contacts (see Fig. \ref{Sup:fig:RandomWalk}a).
The particle starts at the outer contact and then moves on the graphene with a step size of $2.5\,$nm.
Here a move is allowed with equal probability in the up, down, left or right direction.
When the particle reaches the inner or outer contact it will get adsorbed.
The amount of steps, which is proportional to the time diffusing on the graphene, is then recorded.
Additionally, it is observed that $99.8\%$ of the random walks that started at the outer contact also end there.
In Fig. \ref{Sup:fig:RandomWalk}b 1000000 of such random walks are compared and a linear fit is applied to the first 1000 double logarithmic data points.
The linear fit of $ax+b$ yields a slope of $k=-1.5$ which is in accordance with the one-dimensional calculation of Yakimov \cite{Yakimov1980}. According to Ref. \citenum{Yakimov1980}, the exponent for the spectrum is obtained from $k$ as $\gamma=3+k=1.5$.
Further, we investigated the possibility of reflection from the contacts back onto the graphene.
With higher probability of reflection the fitted linear slope decreases to as low as $1.2$ ($\gamma=1.8$) at $75\%$ chance of reflection.

   \section*{IV. Monte Carlo simulations and power spectral density of resistance fluctuations}
Kinetic Monte Carlo simulations provide a powerful tool to investigate 2D dynamics in the presence of particle-particle interactions which are neglected in our basic diffusion calculations.  
In order to elucidate the role of Ne-Ne interactions as the origin of the $1/f^\gamma$ noise observed at $T = 4 - 10$\,K, kinetic Monte Carlo (kMC) simulations were performed on a simplified model system imitating the Corbino geometry \cite{Plimpton2009}, \cite{Sparks}. The simulated system consists of a two-dimensional 50 by 50 square lattice with periodic boundary conditions applied to the left and right boundaries while rigid edges are assumed on the upper and lower boundaries corresponding to the electrodes of an infinitely large Corbino disk. We assume that all lattice sites are equivalent for impurity atoms, \emph{i.e.} we neglect fully ripples on the graphene membrane, although such deformations might lead to clustering of impurities as such. We also neglect the Ne-Ne repulsion at short distances (see Sect. II) and allow the particles to occupy the nearest neighbour sites for computational simplicity. 

In our starting configuration, 25 defects were placed on randomly selected lattice sites. Assuming vacancy diffusion type dynamics \cite{Plimpton2009} - equivalent to the 2-state Ising model performing Kawasaki dynamics, where only one defect can occupy one lattice site at a time and the defects are allowed to move via thermally activated diffusional hops to any of the eight nearest non-defect lattice sites with the rates governed by the following equations:
\begin{subequations}
\begin{align}
  r &= f_0 \exp{(\frac{-E_d}{k_B T})},\; \mathrm{if} \; \Delta E \leq 0 \\
 r &= f_0 \exp{(-\frac{E_d+\Delta E}{k_B T})},\; \mathrm{if} \; \Delta E > 0
\end{align}
 \label{Sub:eq:1}
\end{subequations}
where $r$ is the average rate of a hop to one of the neighboring sites, $f_0$ is the attempt frequency, $\Delta E$ is the change in the system energy and $E_d$ is the activation energy for the diffusional hop, and $T$ is the temperature. $\Delta E$ is determined by the coordination number between the neighboring defects assuming that when the coordination number increases by one, $\Delta E = -2$, and when it decreases by one, $\Delta E = 2$, and so on. Thus, according to Eqs. \ref{Sub:eq:1} the rate of cluster formation is somewhat higher than that of the dissociation and the rates depend on temperature. In our present simulations, simple energy relations were applied also for the other parameters: $f_0=1$, $E_d = 4$ and $k_BT = 1.2$ or $k_BT = 2$.

After the kMC simulations, the produced time series of the positions of the moving defects were used as an input to calculate the corresponding time series of the fluctuating resistance of the system. A minimal model for impurity scattering was employed to estimate the induced resistance change due to the diffusing particles: The resistance of a defect site was taken to be much lower, or alternatively much larger,\footnote{In the experiments, the presence of adsorbed Ne atoms leads to reduced scattering owing to improved screening of Coulomb impurities.}  than that of the background lattice \cite{Lee2016}. In our finite element method (FEM) calculations, \cite{Comsol} we assigned a $10^5$ times smaller (or larger) value for the conductivity at each defect site. Most of our kMC simulations were performed using increased resistance at the defect sites, because atom clusters will act as real scattering centers which eventually win over the screening effects of single atoms.
The resistance at every time step was determined by applying a small DC current from the lower electrode to the upper one and measuring the corresponding voltage. Finally, the power spectral density of the resistance fluctuations (PSD) was calculated and compared to the experimental results.

\begin{figure}[H]
  \centering
  \begin{subfigure}[b]{0.46\linewidth}
    \includegraphics[width=\linewidth]{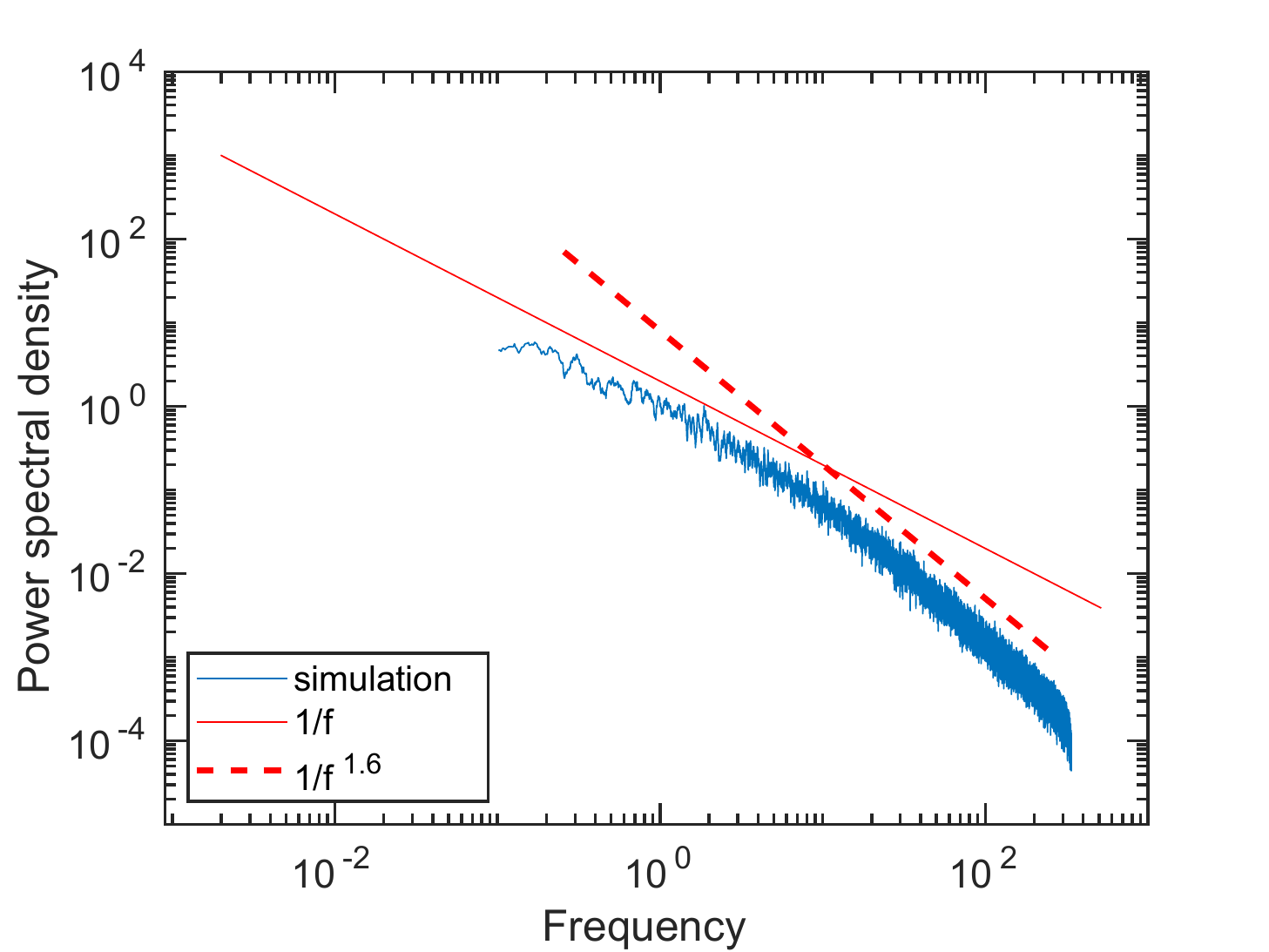}
    \caption{}
  \end{subfigure}
  \begin{subfigure}[b]{0.46\linewidth}
    \includegraphics[width=\linewidth]{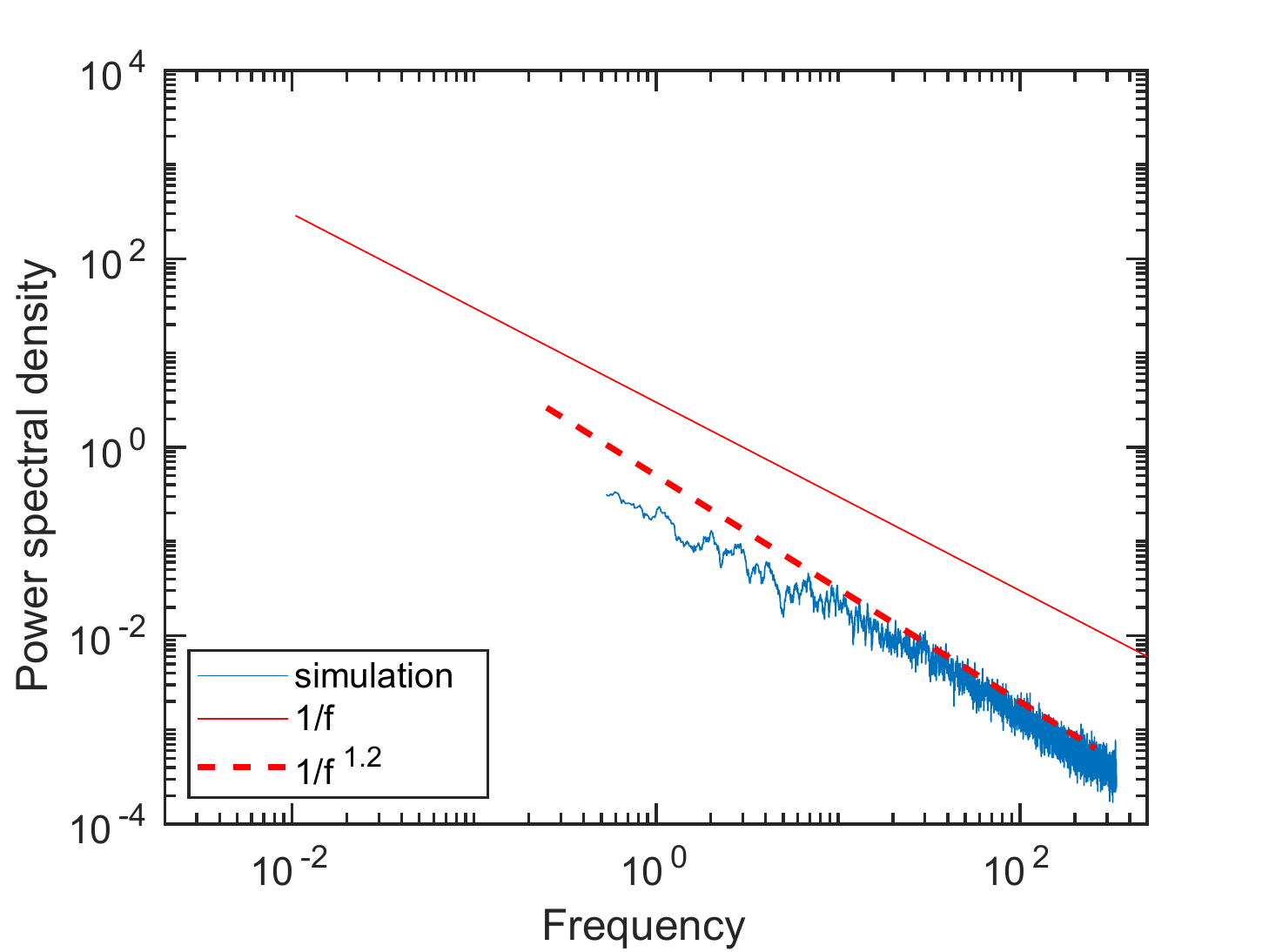}
    \caption{}
  \end{subfigure}
    \caption{Power spectral density of resistance fluctuations for two different temperatures: (a) $k_B T = 1.2$ and (b) $k_B T = 2$. The frequency scale depends on the attempt frequency, here scaled to correspond to $f_0 \sim 10^8$ s$^{-1}$ on the Corbino disk.}
    \label{Sup:fig:PSD}
\end{figure}

\begin{figure}[H]
  \centering
  \begin{subfigure}[b]{0.4\linewidth}
    \includegraphics[width=\linewidth]{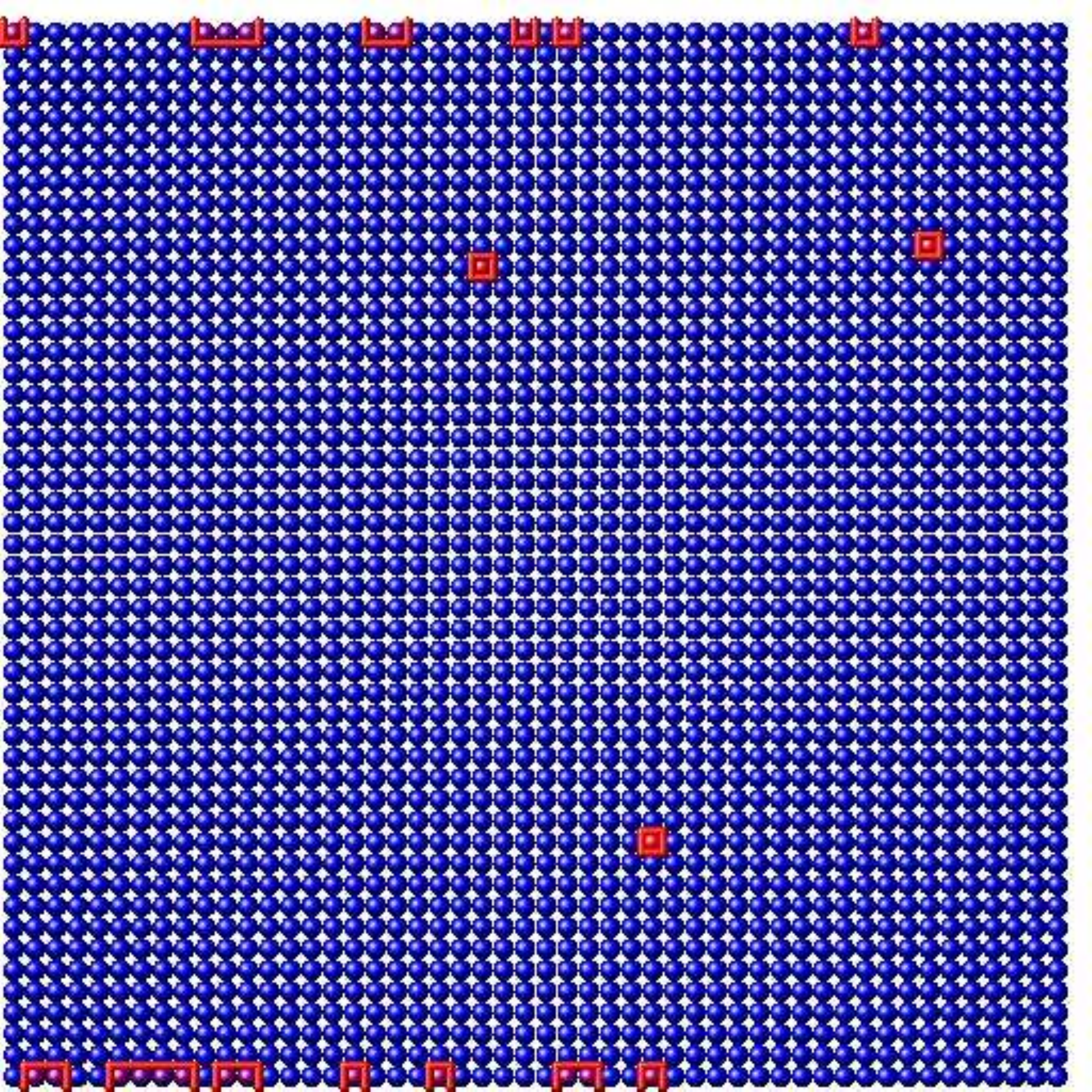}
    \caption{}
  \end{subfigure}
  \begin{subfigure}[b]{0.4\linewidth}
    \includegraphics[width=\linewidth]{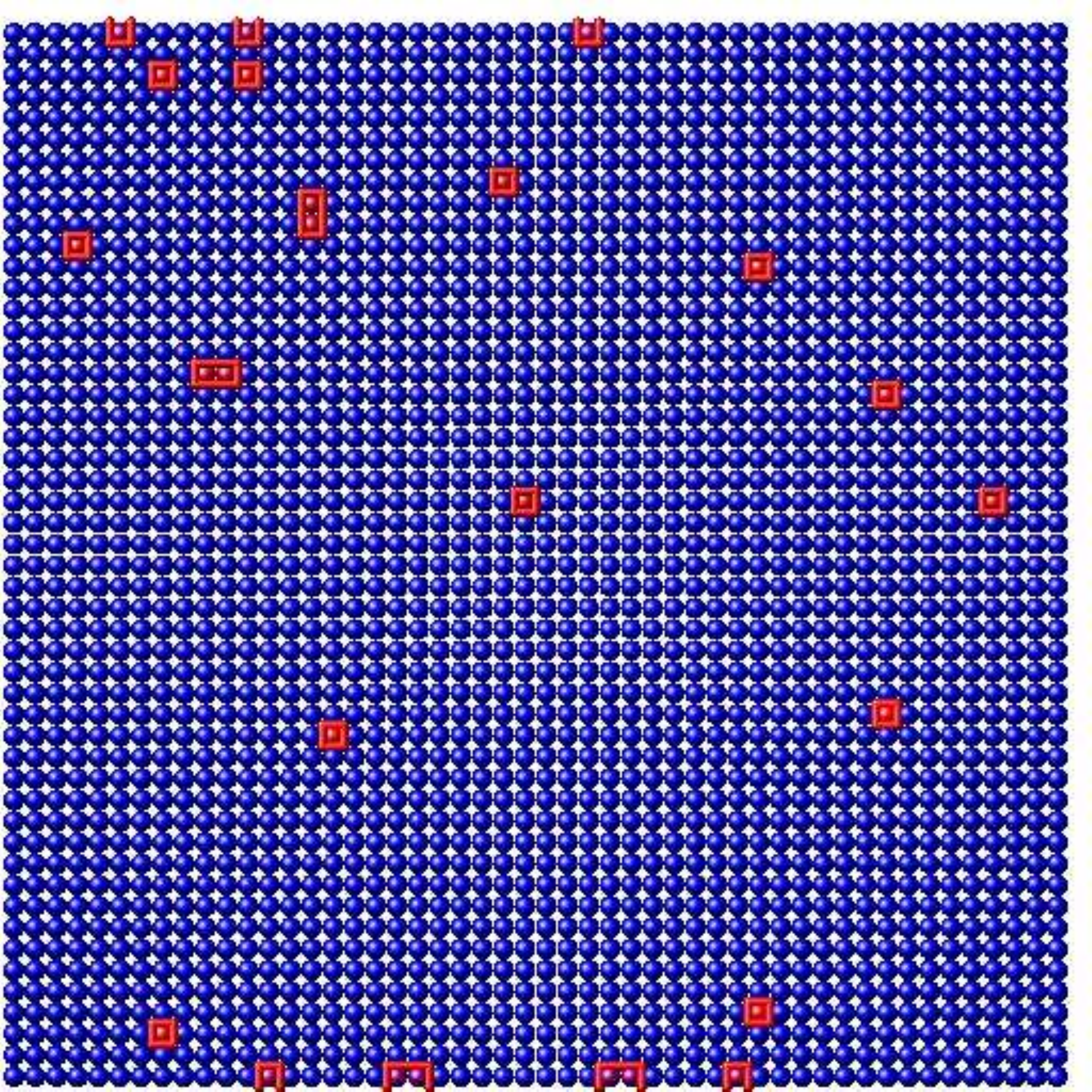}
    \caption{}
  \end{subfigure}
    \caption[Fig7: Snapshots of a trajectory of the defects on the lattice]{Snapshots of a trajectory of the defects on the lattice: (a) $k_B T = 1.2$ and (b) $k_B T = 2$ (for videos see supplementary material mc140\_movie\_small\_dt.mp4 ($k_B T = 1.2$) and mc147\_movie\_small\_dt.mp4 ($k_B T = 2$) ) }
    \label{Sup:fig:defmap}
\end{figure}

Fig. \ref{Sup:fig:PSD} shows the power spectral density of resistance fluctuations for two simulations performed at different temperatures, $k_B T = 1.2$ and $k_B T = 2$, with strong resistance at the scattering sites; practically the same behavior is obtained by setting large conductance at the impurity sites. In the case of $k_B T = 1.2$, the resistance fluctuations follow the power law of $1/f^\gamma$ with $\gamma  \sim 1.6$ near the inspected frequency range, while for the higher temperature the power of $\gamma \sim 1.2$ is exhibited. Based on the kMC simulations, the dynamics of the defects as a function of time was further studied by extracting the average number of defects not in contact with the electrodes showing the average of 3.4 and 13 for $k_B T = 1.2$ and  $k_B T = 2$, respectively. This implies  that the average time the defects spend at the rigid edges, imitating the contact electrodes of the Corbino disk, is significantly longer at the lower temperature, reducing the effective number of mobile defects on graphene. This can also be seen in the attached videos and the representative snapshots of the trajectories shown in Fig. \ref{Sup:fig:defmap}. 

The comparison of Figs. \ref{Sup:fig:PSD} and \ref{Sup:fig:defmap} supports the conjecture that the absorbing boundaries of the electrodes have a significant influence on the observed noise spectra. In more detail, the simulations indicate the importance of cluster formation and dissociation especially to the observed frequency specific power, $\gamma$, of $1/f^\gamma$: the average number of mobile clusters is more than four times higher for $k_B T = 2$, corresponding to $\gamma \sim 1.2$,  as compared to the case of $k_B T = 1.2$, where single mobile defects dominate the dynamics at the inspected high frequency range, corresponding to $\gamma \sim 1.6$. At lower frequencies where longer correlations become visible, the PSD curves approach the form of $1/f$ noise. In the case of $k_B T = 2$ (Figs. \ref{Sup:fig:PSD}(b) and \ref{Sup:fig:defmap}(b)), the clustering based correlations are visible also at higher frequencies due to the larger number of mobile clusters.

\section*{V. Graphene noise vs. contact noise}
In the main paper, we considered the $1/f^{\gamma}$ noise as coming from graphene without trying to separate the exact origin of the noise, whether it comes purely from graphene or whether it is also related to electrical contacts. As is well known, the resistance in high quality graphene samples originates mostly from the contacts, and the same could happen with the $1/f^{\gamma}$ noise. The separation of noise contribution from contacts has been discussed in clean graphene in Ref. \citenum{Masahiro2021} in presence of incoherent noise sources. Specific features of the measured $S_I(V_g)$ could be related to contact noise $S_I^c$ and to the graphene noise $S_I^{gr}$. In particular, a M-shaped $S_I(V_g)$ curve with leveling off at large charge densities could be explained using the incoherent noise source model, in which the dip in the noise is related to $S_I^c$  and to the coexistence of electrons and holes \cite{Masahiro2021}. Moreover, the contact noise governs the leveling off of $S_I(V_g)$ at large carrier densities.

\begin{figure}[H]
    \centering
    \includegraphics[width=0.8\linewidth]{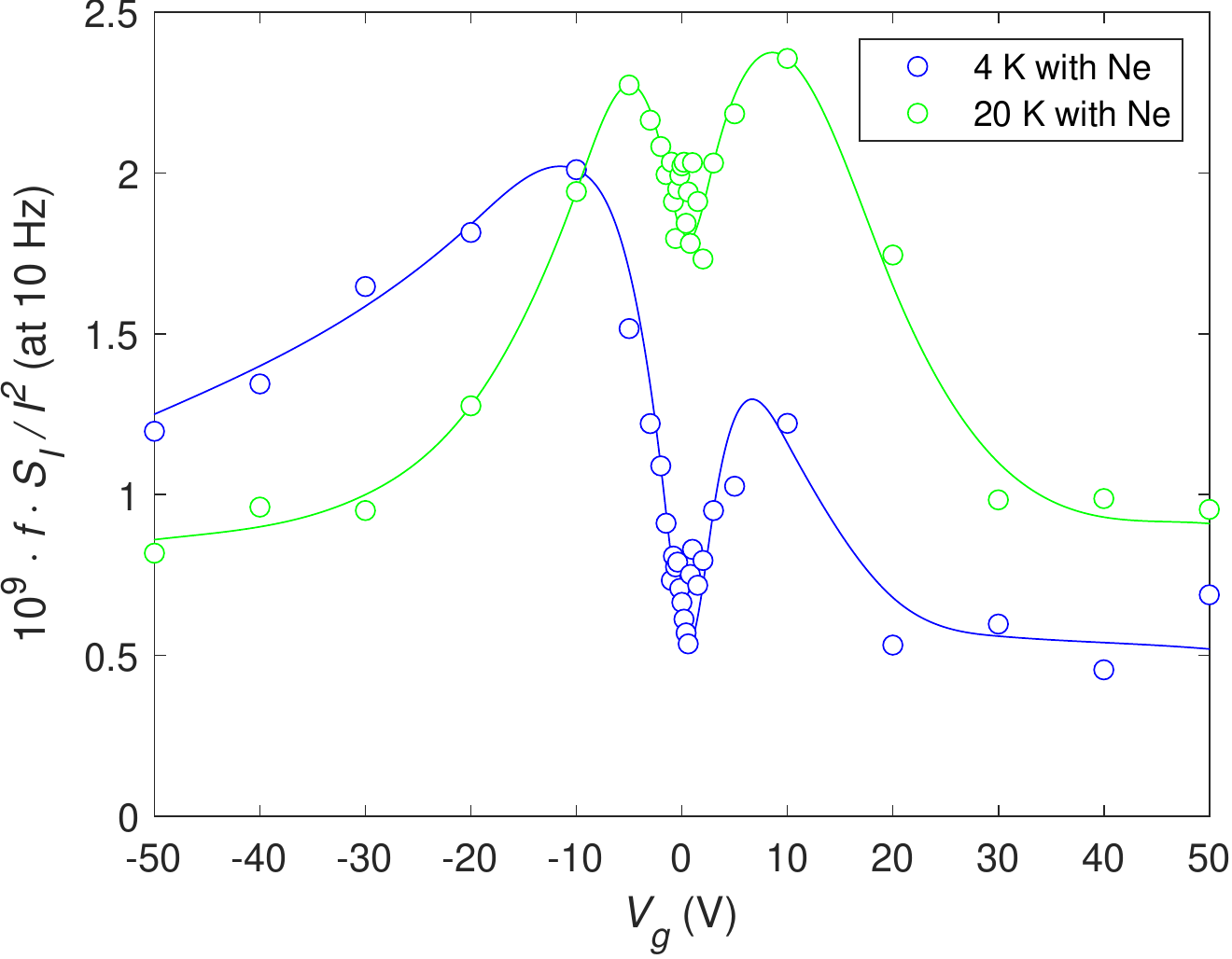}
    \caption{Gate voltage dependence of scaled current noise $f \cdot S_I/I^2$ at 10 Hz measured with adsorbed Ne atoms at $T = 4$\,K (blue) and $T = 20$\,K (green) . The overlaid curves are to guide the eyes.}
    \label{Sup:fig:NoiseVSgate}
\end{figure}

Fig. \ref{Sup:fig:NoiseVSgate} displays current noise $S_I(V_g)$ data measured with adsorbed Ne atoms at $T = 4$\,K and $T = 20$\,K. The data display a well-defined minimum of noise at the Dirac point, followed by a maximum in $S_I(V_g)$  at gate voltage $V_g \simeq \pm 10$\,V, and finally a clear decrease of noise at $|V_g| > 10$\,V which tends to saturation when $|V_g| \rightarrow 50$\,V. The behavior thus follows exactly the $V_g$ dependence outlined in the analysis of Ref. \citenum{Masahiro2021}. Consequently, we can conclude that the division between contact noise $S_I^c$ and graphene noise $S_I^{gr}$ is qualitatively similar in the presence of adsorbed Ne as in clean graphene. From the data at $T = 4$\,K, however, we may infer that the contact noise becomes asymmetric with respect to electrons and holes: electron conduction is seen to have smaller noise than the hole conduction at $V_g < 0$. Comparing with Fig. \ref{SI:fig:RvsVg}, we see that the change in contact resistance due to Ne atoms is stronger in the hole carrier regime, \emph{i.e.} under similar conditions as $S_I^c$. Hence, we conclude that both contact noise and graphene noise are influenced by adsorbed Ne. Apart from local pseudomagnetic fields due to individual adsorbed atoms, the noise in both cases is due to changes in the local doping of graphene due to variation in Ne-induced scalar potential. Our kMC analysis with Corbino like boundary conditions includes significant clustering of atoms at the contacts and thereby noise due to changes in the boundary layer are an integral part of our numerical noise analysis. On broader scale, the origin of the contact noise is the same as that of graphene, even though the presence of the Au side wall changes the basic conditions for clustering.

%\bibliography{CorbinoMgNoiseRefs}
%\bibliographystyle{plain}
\providecommand{\latin}[1]{#1}
\makeatletter
\providecommand{\doi}
  {\begingroup\let\do\@makeother\dospecials
  \catcode`\{=1 \catcode`\}=2 \doi@aux}
\providecommand{\doi@aux}[1]{\endgroup\texttt{#1}}
\makeatother
\providecommand*\mcitethebibliography{\thebibliography}
\csname @ifundefined\endcsname{endmcitethebibliography}
  {\let\endmcitethebibliography\endthebibliography}{}

\end{document}